\documentclass[12pt]{iopart} 
\pdfoutput=1 
\usepackage{epsfig,psfrag}
\usepackage{amsmath,amsfonts,amssymb}
\usepackage{bbm}
\usepackage{mathrsfs}

\usepackage{graphicx}
\usepackage[usenames]{color}
\usepackage{amsbsy}
\newcommand{\cc}{{\mbox{c.c.\,}}}
\newcommand{\cZ}{\mathcal Z}
\newcommand{\cU}{\mathcal U}

\def\tr{\,{\rm tr}\,}
\newcommand{\fract}[2]{{\textstyle\frac{#1}{#2}}}

 \newcommand{\p}{\partial}

\newcommand{\be}{\begin{equation}}
\newcommand{\ee}{\end{equation}}
\newcommand{\bea}{\begin{eqnarray}}
\newcommand{\eea}{\end{eqnarray}}
\newcommand{\ben}{\begin{enumerate}}
\newcommand{\een}{\end{enumerate}}
\newcommand{\bit}{\begin{itemize}}
\newcommand{\eit}{\end{itemize}}

\newcommand{\Eq}[1]{Eq.~(\ref{#1})}
\numberwithin{equation}{section}

\def\a{\alpha}
\def\b{\beta}

\def\k{\kappa}
\newcommand{\cE}{\ensuremath{{\cal E}}}
\definecolor{BrickRed}{cmyk}{0,0.89,0.94,0.28}
\definecolor{MidnightBlue}{cmyk}{0.98,0.13,0,0.43}
\definecolor{DarkGreen}{rgb}{0,0.7,0.1}

\newcommand{\bfx}{{\bf x}}

\newcommand{\cC}{{\mathcal C}}
\newcommand{\cF}{{\mathcal F}}
\newcommand{\cEC}{{\mathcal E}[{\mathcal C}]} 

\newcommand{\cD}{{\mathcal D}}
\newcommand{\cT}{{\mathcal T}}
\newcommand{\cG}{{\mathcal G}}
\newcommand{\cO}{{\mathcal O}}

\newcommand{\phicl}{{\ensuremath{\phi_{\rm cl}}}}
\newcommand{\bphi}{\mbox{\boldmath$\phi$}}

\newcommand{\bbS}{\mathbb S}
\newcommand{\bbU}{\mathbb U}
\newcommand{\bbT}{\mathbb T}
\newcommand{\bbI}{\mathbb I}

\newcommand{\bbN}{\mathbb N}

\newcommand{\hr}{\hat{\mathbf x}}
\newcommand{\hz}{\hat{\mathbf z}}

\newcommand{\bA}{\mathbf A}
\newcommand{\bE}{\mathbf E}
\newcommand{\bB}{\mathbf B}
\newcommand{\bD}{\mathbf D}
\newcommand{\bJ}{\mathbf J}
\newcommand{\bH}{\mathbf H}
\newcommand{\bM}{\mathbf M}

\newcommand{\bQ}{\mathbf Q}
\newcommand{\bP}{\mathbf P}

\newcommand{\bpsi}{\mbox{\boldmath$\Psi$}}

\begin{document} 
\title{
Casimir forces between arbitrary compact objects: Scalar and
electromagnetic field}

\author{T. Emig$^{1,2}$
and
R. L. Jaffe$^{3}$
}

\address{$^1$ Institut f\"ur Theoretische Physik, Universit\"at zu
  K\"oln, Z\"ulpicher Strasse 77, \\50937 K\"oln, Germany}
\address{$^2$ Laboratoire de Physique Th\'eorique et Mod\`eles
  Statistiques, CNRS UMR 8626, B\^at.~100, Universit\'e Paris-Sud, 91405
  Orsay cedex, France}
\address{$^3$ Center for Theoretical Physics, Laboratory for Nuclear
  Science, and Department of Physics, Massachusetts Institute of
  Technology, Cambridge, MA 02139, USA}

\begin{abstract} 
  We develop an exact method for computing the Casimir energy between
  arbitrary compact objects, both with boundary conditions for a
  scalar field and dielectrics or perfect conductors for the
  electromagnetic field. The energy is obtained as an interaction
  between multipoles, generated by quantum source or current
  fluctuations.  The objects' shape and composition enter only through
  their scattering matrices.  The result is exact when all multipoles
  are included, and converges rapidly.  A low frequency expansion
  yields the energy as a series in the ratio of the objects' size to
  their separation.  As examples, we obtain this series for two
  spheres with Robin boundary conditions for a scalar field and
  dielectric spheres for the electromagnetic field. The full
  interaction at all separations is obtained for spheres with Robin
  boundary conditions and for perfectly conducting spheres.
\end{abstract} 
\submitto{\JPA} 

\section{Introduction}
\label{introduction}

Casimir forces arise when the quantum fluctuations of a scalar,
vector, or even fermion field are modified by the presence of static
or slowly changing external objects~\cite{Casimir:1948dh}.  The
objects can be modeled by boundary conditions that they place on {the
  fluctuating field $\phi$}, by an external field, $\sigma$, to which
$\phi$ couples~\cite{sigmawork}, or, in the case of electromagnetism,
by a material with space and frequency dependent dielectric and
magnetic properties.  The Casimir energy is the difference between the
energy of the fluctuating field {when then objects are present} and
when the objects are removed to infinite separation.

The advent of precision experimental measurements of Casimir forces
\cite{Bordag:2001qi} and the possibility that they can be applied to
nanoscale electromechanical devices~\cite{chan,Capasso:2007nq} has
stimulated interest in developing a practical way to calculate the
dependence of Casimir energies on the shapes of the objects.  Many
geometries have been analyzed over the years, but the case of compact
objects has proved rather difficult.  In a recent Letter we described
a new method that makes possible accurate and efficient calculations
of Casimir forces and torques between any number of compact objects
\footnote{This presentation is based on work performed in
  collaboration with N.~Graham and M.~Kardar.  For a complete
  exposition, see Refs.~\cite{Emig:2007cf} and \cite{Emig+07}.}.  The method applies
to electromagnetic fields and dielectrics as well as perfect
conductors.  It {also} applies to other fields, {such as} scalar and
Dirac, and to any boundary conditions.  {In this approach,} the
Casimir energy is given in terms of the {fluctuating field's}
scattering amplitudes from the individual objects, {which encode} the
effects of the shape and boundary conditions.  The scattering
amplitudes are known analytically in some cases and numerically in
others.  If the scattering amplitudes are known, then the method can
be applied from asymptotically large separation down to separations
that are a small fraction of the dimension of the objects.  Results at
large separations are obtained using low frequency and low angular
momentum {expansions of} scattering amplitudes.  The coefficients
multiplying the successive orders in inverse separation can be
identified with increasingly detailed characteristics of the objects.
At small separations the manipulation of large matrices, whose
dimensions grow with angular momentum, eventually slows down the
calculation.  However at these distances other methods, notably the
``proximity force approximation'' {(PFA)}, apply.  Thus it is now
possible to obtain an understanding of Casimir forces and torques at
all separations for compact objects.

The aim of this talk is to provide a pedagogical introduction to our
methods by treating in detail the simplest case, a scalar field
obeying a boundary condition on a sharp surface \cite{Emig+07}.  The complications of
electromagnetism and smoothly varying dielectrics were already
introduced in {Ref.}~\cite{Emig:2007cf} and are discussed here briefly. 
Our result for the Casimir energy is remarkably simple.  For a complex
scalar field in the presence of two objects it takes the form,
\begin{equation}
\label{tform}
\cE_{12}[\cC]=\frac{\hbar c}{\pi}\int_0^\infty d\kappa \,
\ln\det(\mathbb{I}-\mathbb{T}^{1}\mathbb{U}^{12}
\mathbb{T}^{2}\mathbb{U}^{21}) \, ,
\end{equation}
where the determinant is over the partial wave indices on transition
matrices $\mathbb{T}^\alpha$ and translation matrices
$\mathbb{U}^{\alpha\beta}$ and the integral is over
$\kappa=-i\omega/c$, the imaginary wavenumber.  For a (real)
electromagnetic field the result has the same form with an additional
factor of $1/2$ and the appropriate matrices for scattering and
propagation of electromagnetic waves.  In Sections \ref{appl} and
\ref{sec:appl-em} the usefulness of this result is demonstrated
through several specific applications for scalar and electromagnetic
fields, respectively.

The force between atoms at asymptotically large distances was computed
by Casimir and Polder~ \cite{Casimir+48} and related to the atoms'
polarizabilities.  For compact objects, such as two spheres, Feinberg
and Sucher~\cite {Feinberg+70} generalized this work to include
magnetic effects. Earlier studies of the Casimir force between compact
objects include a multiple reflection formalism~\cite{Balian}, which in
principle could be applied to perfect conductors of arbitrary shape.
A formulation of the Casimir energy of compact objects in terms of
their scattering matrices, for a scalar field coupled to a dielectric
background, is introduced in Ref.~\cite{Kenneth+06}, where it is
suggested that it can also be extended to the EM case.

Recently Gies et al.~\cite{Gies:2003cv} used numerical methods to
evaluate the Casimir force between two Dirichlet spheres for a scalar
field, over a range of subasymptotic separations{, and in other open
  geometries {such} as a plate and a cylinder \cite{Gies+06b} or
  finite plates with edges \cite{Gies+06a}.}  Bulgac and
collaborators~\cite{Bulgac:2005ku} applied scattering theory methods
to the same {scalar Dirichlet} problem and obtained results over a
wide range of separations.  The only explicit calculations for
subasymptotic distances up to now have been for a scalar field obeying
Dirichlet boundary conditions on two spheres, a sphere and a
plate~\cite{Bulgac:2005ku} and for electromagnetic fields for a plate
and a cylinder \cite{Emig+06a} and two perfectly conducting spheres
\cite{Emig:2007cf}.

\section{Casimir energy from functional integral: Scalar field}

In this Section we review formalism essential for our work that is
based on the functional integral approach of
Refs.~\cite{LK91,Emig+01,Buscher+05}.

\subsection{Functional integral formulation}
\label{2.1}
We consider a complex quantum field, $\phi(\bfx,t)$, which is defined
over all space {and} constrained by boundary conditions
{$\cal C$} on a set of fixed surfaces $\Sigma_\alpha$, for
$\alpha=1,2, \dots, N$, but is otherwise non-interacting.  We assume that the
surfaces are closed and compact  and refer to their interiors as
``objects.''  Our starting point is the functional integral
representation for the trace of the propagator, ${\rm Tr}\,
e^{-iH_{\cC}T/\hbar}$~\cite{fandh},
\begin{equation}
\label{effact}
{{\rm Tr}\, e^{-iH_{\cC}T/\hbar}=\int \left[{\cal D}\phi\right]_{\cal C}  
\, e^{\frac{i}{\hbar}S_{E}[\phi]}\,\equiv\, Z\,[\cC] \, ,}
\end{equation}
where the subscript $\cC$ denotes the constraints imposed by the
boundary conditions.\footnote{We have used an abbreviated
  notation for the functional integral.  Since $\phi$ is complex
  $\int\cD\phi$ should be understood as $\int\cD\phi\cD\phi^{*}$, and
  similarly in subsequent functional integrals.}
The integral is over all field {configurations} that obey
the boundary conditions and are periodic in a time interval $T$.
$S[\phi]$ is the action for a free complex field,
\begin{equation}
\label{action1}
S_{E}[\phi]= \int_{0}^{T}dt\int d\bfx\,
\left({\frac{1}{c^2}}|\partial_t\phi|^{2}-|{\nabla}\phi|^{2}\right) ,
\end{equation}
where the $\bfx$-integration covers all space.\footnote{Note that
  $\phi$ is defined and can fluctuate inside the objects bounded by
  the surfaces $\Sigma_\alpha$.  In this feature our formalism departs
  from some treatments where the field is defined to be strictly zero
  (for Dirichlet boundary conditions) inside the objects.  The
  fluctuations interior to the objects do not depend on the
  separations between them and therefore do not affect Casimir forces
  or torques.}
 
The ground state energy can be projected out of the trace in 
Eq.~\eqref{effact} by setting $T=-i\Lambda/c$ taking the limit
$\Lambda\to\infty$,
\begin{equation}
\label{proj}
{\cE}_{0}[\cC] = -\lim_{\Lambda\to\infty}\frac{\hbar c}{\Lambda}
\ln\left({\rm Tr \,} 
e^{-H_{\cC}\Lambda/\hbar c}\right)=-\lim_{T\to\infty}
\frac{\hbar c}{\Lambda} \ln Z[\cC] \, ,
\end{equation}
and the Casimir energy is obtained by subtracting the ground state
energy when the objects have been removed to infinite separation,
\begin{equation}
\label{casimir1}
\cEC = -\lim_{\Lambda\to\infty}\frac{\hbar c}{\Lambda}
\ln\left( Z[\cC]/ Z_{\infty}\right) \, .
\end{equation}
In the standard formulation, the constraints are implemented by
boundary conditions on the field $\phi$ at the surfaces
$\{\Sigma_\alpha\}$.  The usual choices are Dirichlet, $\phi=0$,
Neumann, $\p_{n}\phi=0$, or mixed {(Robin)}, $\phi {-}
\lambda\p_{n}\phi=0$, where $\p_{n}$ is the normal derivative pointing
out of the objects.  To be specific, we first consider Dirichlet
boundary conditions.  The extension to the Neumann case is presented
in Ref~\cite{Emig+07}.  As noted in the Introduction, the only effect
of the choice of boundary conditions is to determine which
$\mathbb{T}$-matrix appears in the functional determinant,
Eq.~\eqref{tform}.

Since the constraints on $\phi$ are time independent, the integral
over $\phi(\bfx, t)$ may be written as an infinite product of
integrals over Fourier components,
\begin{equation}
\int \left[{\cal D}\phi\right]_{\cal C} =
\prod_{n=-\infty}^{\infty}\left[{\cD}\phi_{n}(\bfx)\right]_{\cC} \, ,
\end{equation}
where 
\begin{equation}
\phi(\bfx, t)=\sum_{n=-\infty}^{\infty}\phi_{n}(\bfx)e^{2\pi i n t/T} ,
\end{equation}
and the logarithm  of $Z$ becomes a sum, 
\begin{equation}
\label{firstlog}
\ln Z[\cC] = \sum_{n=-\infty}^{\infty}
\ln\left\{\int\left[{\cD}\phi_{n}(\bfx)\right]_{\cC} 
\exp \left[ i\frac{T}{\hbar}\int d\bfx
\left(\left(\frac{2\pi n}{{c}T}\right)^{2}|
\phi_{n}(\bfx)|^{2}- |\nabla\phi_{n}(\bfx)|^{2}  \right)\right]\right\} \, .
\end{equation}

As $T\to\infty$,  $\sum_n$ can be replaced by
$\fract{cT}{2\pi}\int_{-\infty}^{\infty}dk$, 
where $k=2\pi n /(cT)$ and $\phi_{n}(\bfx)$
is replaced by $\phi(\bfx,k)$.  Combining the positive and negative
$k$-integrals gives
\begin{eqnarray}
\label{secondlog}
\hspace*{-1.3cm}
\ln Z[\cC] 
&=&  \frac{cT}{\pi}\int_{0}^{\infty}d  k\ln 
\left\{\int\left[{\cD}\phi (\bfx,  k)\right]_{\cC} 
\exp \left[ i\frac{T}{\hbar}\int d\bfx\left(  k^{2}|\phi(\bfx,  k)|^{2} -
|\nabla\phi(\bfx,  k)|^{2}\right)\right]\right\}
\nonumber\\
&=&\frac{cT}{\pi}\int_{0}^{\infty}d  k\ln\, \mathfrak{Z}_{\cC}(  k) \, ,
\end{eqnarray}
where
\begin{equation}
\label{minkowski-z}
\mathfrak{Z}_{\cC}( k) =\int\left[{\cD}\phi (\bfx,k)\right]_{\cC} 
\exp \left[ i\frac{T}{\hbar} \int d\bfx\left(k^{2}|\phi(\bfx,k)|^{2} -
|\nabla\phi(\bfx,k)|^{2}\right)\right],
\end{equation}
is the functional integral at fixed $k$.

To extract the Casimir energy, we use $T=-i\Lambda/c$ and Wick rotate
the $k$-integration ($k=i\kappa$ with $\kappa>0$){.}
\footnote{A more
careful treatment of the rotation of the integration contour to the
imaginary axis is necessary in the presence of bound states.}  Using
Eq.~\eqref{casimir1}, we obtain,
\begin{equation}
\label{casimir2}
\cEC = {-} \frac{\hbar c}{\pi}\int_{0}^{\infty} d\kappa
\ln\frac{\mathfrak{Z}_{C}(i\kappa)}{\mathfrak{Z}_{\infty}(i\kappa)} \, .
\end{equation}
{Here} $\mathfrak{Z}_{\cC}(i\k)$ is given by the Euclidean
functional integral,
\be
\label{euclid-z}
\mathfrak{Z}_{\cC}(i\kappa) =\int\left[{\cD}\phi (\bfx,i\k)\right]_{\cC} 
\exp \left[ - \frac{T}{\hbar} \int d\bfx\left(\k^{2}|\phi(\bfx,i\k)|^{2}
+ |\nabla\phi(\bfx,i\k)|^{2}\right)\right]\, .
\ee

It remains to incorporate the constraints directly into the functional
integral using the methods of Refs.~\cite{Bordag+85,LK91}.  Working in Minkowski
space, we consider the fixed frequency functional
integral, $\mathfrak{Z}_{\cC}(k)$ (and suppress the label $k$ on the
field $\phi$).  Following Ref.~\cite{Bordag+85,LK91}, we implement the
constraints in the functional integral by means of a functional
$\delta$-function.   For Dirichlet boundary conditions
the constraint reads,
\begin{equation}
\label{deltafn}
\int \left[{\cD}\phi (\bfx)\right]_{\cC}=\int \left[{\cD}\phi (\bfx)\right] 
\prod_{\a=1}^{N}\int \left[{\cD}\varrho_{\a}(\bfx)\right]
\exp\left[ i \frac{T}{\hbar}\int_{\Sigma_{\a}} 
d\bfx \left(\varrho^{*}_{\a}(\bfx)\phi(\bfx)+ \cc\right)\right] \, ,
\end{equation}
where the functional integration over  $\phi$ is no longer
constrained.  Other boundary conditions can be
implemented similarly.  In the resulting functional integral,
\begin{eqnarray}
\hspace*{-1cm}
\mathfrak{Z}_{\cC}(k)&=&\prod_{\a=1}^{N}\int
\left[{\cD}\varrho_{\a}(\bfx )
\right]\int \left[{\cD}\phi 
(\bfx)\right] \exp \left[ i\frac{T}{\hbar}\left( 
\int d\bfx\left(k^{2}|\phi(\bfx)|^{2}-|
\nabla\phi(\bfx)|^{2}  \right) \right. \right. \nonumber\\
 &+& \left. \left. \sum_{\a}\int_{\Sigma_{\a}} d\bfx
\left(\varrho^{*}_{\a}(\bfx )
\phi(\bfx )+ \cc\right)\right)\right]\nonumber\\
&\equiv&\prod_{\a=1}^{N}\int \left[{\cD}\varrho_{\a}(\bfx )\right]\int 
\left[{\cD}\phi (\bfx)\right] 
\exp \left(i\frac{T}{\hbar}\widetilde S[\phi,\varrho]\right) \, ,
\label{source3}
\end{eqnarray} 
the fields fluctuate without constraint throughout space and the
sources $\{\varrho_{\a}\}$ fluctuate on the surfaces.  We denote the new
``effective action'' including both the fields and sources by
$\widetilde S[\phi,\varrho]$.

\subsection{Performing the integral over $\phi$}

We start with the expression for the fixed-$k$ functional integral,
\Eq{source3}.  For any fixed sources, $\{\varrho_{\a}\}$, there is a
unique classical field, $\phicl[\varrho]$, that is the solution to
$\delta \widetilde S[\phi,\varrho]/\delta\phi(\bfx)=0$.  The classical
theory defined by $\widetilde S[\phi,\varrho]$, describes a complex
scalar field coupled to a set of sources on the surfaces, and is a
generalization of electrostatics.  By analogy with electrostatics, the
field $\phi$ is continuous throughout space, but its normal derivative
jumps by $\varrho_{\a}(\bfx )$ across $\Sigma_{a}$.  Indeed, the
classical equations of motion that follow from $\delta\widetilde
S/\delta\phi=0$ are
\begin{eqnarray}
\left(\nabla^{2} +k^{2}\right)\phicl(\bfx)&=& 0,
\quad\mbox{for $\bfx\notin \Sigma_\a$},\nonumber\\
\Delta \phicl(\bfx )&=& 0,
\quad\mbox{for $\bfx\in \Sigma_\a$,}\nonumber\\
\left.\Delta \p_{n}\phicl\right|_{\bfx }&=&\varrho_{\a}(\bfx ),
\quad \mbox{for $\bfx\in \Sigma_\a$,}
\label{variation}
\end{eqnarray}
where $\Delta\phi=\phi_{\rm in}-\phi_{\rm out}$ and
$\Delta\p_{n}\phi=\p_{n}\phi|_{\rm in}-\p_{n}\phi_{\rm out}$. The
subscripts ``in'' and ``out'' refer to the field inside and outside
the bounding surface $\Sigma_\a$.  As before, all normals point out of
the compact surfaces.  The solution to Eq.~\eqref{variation} is unique
up to solutions of the homogeneous equations, which we
exclude by demanding that $\phicl$ vanish when the
$\{\varrho_{\alpha}\}=0$.  Continuing the analogy with electrostatics,
we can write the classical field in terms of the free Green's function
and the sources,
\begin{equation}
\label{phib}
\phicl(\bfx) =\sum_{\beta}\int_{\Sigma_{\b}}d\bfx'
\cG_{0}(\bfx,\bfx',k)\varrho_{\b}(\bfx') \, ,
\end{equation}
where the free Green's function is given by
\begin{eqnarray}
\label{gf3}
{\cal G}_{0}(\bfx,\bfx',k) &\equiv& \frac{e^{ik|\bfx-\bfx'|}}{4\pi|\bfx-\bfx'|}
= {ik}\sum_{lm}j_{l}(k r_{<})
h^{(1)}_{l}(k r_{>})Y_{lm}(\hat\bfx)Y_{lm}^{*}(\hat\bfx') \nonumber\\
&=&{ik}\sum_{lm}j_{l}(k r_{<})
h^{(1)}_{l}(k r_{>})Y_{lm}(\hat\bfx')Y_{lm}^{*}(\hat\bfx) \, ,
\end{eqnarray}
where the notations $r_{<(>)}$ refer to whichever of $r,r'$ is the
smaller (larger).  

To compute the functional integral over $\phi$, {we} first
decompose $\phi$ into the classical part given by Eq.~\eqref{phib} and
a fluctuating part, 
\begin{equation}
\phi(\bfx) = \phicl(\bfx) +\delta \phi(\bfx) \, .
\end{equation}
Then, because the effective action, $\widetilde S$, is quadratic in
$\phi$, the $\delta\phi$ dependent terms are independent of $\phicl$,
\begin{equation}
\label{sourceintegral1}
\mathfrak{Z}_{\cC}(k)= \prod_{\a=1}^{N}
\int \left[{\cD}\varrho_{\a}(\bfx )\right]
e^{i {\frac{T}{\hbar}}
\widetilde S_{\rm cl}[\varrho]}\int [{\cD}\delta\phi (\bfx)]
\exp \left[i  {\frac{T}{\hbar}} \int
d\bfx\left(k^{2}|\delta\phi(\bfx)|^{2}- |\nabla\delta\phi(\bfx)|^{2}  
\right)\right] \, .
\end{equation}
The classical action can be simplified by using the equations of
motion, Eq.~\eqref{variation}, which make it possible to express the action
entirely in terms of integrals over the surfaces $\{\Sigma_{\a}\}$,
\begin{equation}
\label{classicalaction}
\widetilde S_{\rm cl}[\varrho] = {\frac{1}{2}}\sum_{\a}
\int_{\Sigma_{\a}} d\bfx \,
\left(\varrho^{*}_{\a}(\bfx )\phicl(\bfx )+\cc \right)\, ,
\end{equation}
where $\phicl(\bfx )$ is understood  to be a functional of the sources
{$\varrho_{\a}$}. 

The functional integral over $\delta\phi$ is \emph{independent of the
classical field $\phicl$} and defines the energy of the
unconstrained vacuum fluctuations of $\phi$.  This term is divergent, or,
more precisely, depends on some unspecified ultraviolet cutoff.
However it can be discarded because it is independent of the sources
and therefore common to $\mathfrak{Z}_{\cC}$ and $\mathfrak{Z}_{\infty}$.
Note that this result is an explicit demonstration of the contention of
Ref.~\cite{rlj}: the Casimir force has nothing to do with the
\emph{vacuum} fluctuations of $\phi$, but is instead a consequence of
the interaction between fluctuating sources in the materials.  It is
therefore not directly relevant to the fluctuations that are conjectured to be
associated with the dark energy.

From Eq.~\eqref{phib} it is clear that the solution to
Eq.~\eqref{variation} obeys the superposition principle:
$\phicl(\bfx)$ is a sum of contributions from each of the sources,
\begin{equation}
\phicl(\bfx)=\sum_{\beta}\phi_{\beta}(\bfx) \, ,
\end{equation}
where $\phi_{\beta}$ satisfies Eq.~\eqref{variation} with all sources
set equal to zero except for $\varrho_{\beta}$.  So the action can be
expressed as a double sum over surfaces and over contributions to
$\phicl$ generated by different objects.  This leaves a partition
function, $\mathfrak{Z}_{\cC}(k)$, of the form
\begin{equation}
\label{jintegral}
\mathfrak{Z}_{\cC}(k)= \prod_{\a=1}^{N}
\int \left[{\cD}\varrho_{\a}(\bfx )\right]
\exp\left[{\frac{i}{2} {\frac{T}{\hbar}} }\sum_{\a,\b}
\int_{\Sigma_\a} d\bfx \left(\varrho^{*}_{\a}(\bfx )
\phi_{\beta}(\bfx )+\cc\right)\right],
\end{equation}
to be evaluated.

\subsection{Evaluation of the Classical Action}

The classical action in Eq.~\eqref{jintegral} contains two
qualitatively different terms, the interaction between different
sources, $\a\ne\beta$, and the self-interaction of the source
$\varrho_{\a}$.  Both can be expressed as functions of the multipole
moments of the sources on the surfaces.

\subsubsection{Interaction terms:  $\a\ne\beta$}

Consider the contribution to the action from the field,
$\phi_{\beta}$, generated by the source, $\varrho_{\beta}$, integrated over
the surface $\Sigma_{\alpha}$,
\begin{equation}
\label{interaction}
\widetilde S_{\b\a}=\frac{1}{2}\int_{\Sigma_\a} d\bfx_\alpha  \left(
\varrho^{*}_{\a}(\bfx_\alpha )\phi_{\beta}(\bfx_\alpha)+\cc \right) \, ,
\end{equation}
where the subscript $\alpha$ on $\bfx_\alpha$ indicates that the
integration runs over coordinates measured relative to the origin of
object $\alpha$.  The field $\phi_{\b}(\bfx_\beta)$, \emph{measured
relative to the origin of object $\beta$}, can be represented as an
integral over its sources on the surface $\Sigma_{\b}$ as in
Eq.~\eqref{phib}.  Since every point on $\Sigma_{\alpha}$ is outside
of a sphere enclosing $\Sigma_{\b}$, the partial wave representation
of $\cG_{0}$ simplifies.  The coordinate $\bfx_{>}$ is always
associated with $\bfx_\beta$ and $\bfx_{<}$ is identified with
$\bfx'_\beta$, so Eq.~\eqref{phib} can be written
\begin{equation}
\label{phib2}
\phi_{\b}(\bfx_\beta)= {ik}\sum_{lm} h^{(1)}_{l}(k r_{\b})Y_{lm}(\hat\bfx_{\b})
\int_{\Sigma_{\b}}d\bfx'_{\b}j_{l}(k r'_{\b})
Y^{*}_{lm}(\hat\bfx'_{\b})\varrho_{\b}(\bfx'_{\b})\, .
\end{equation}
Note that the arguments of the Bessel functions and spherical
harmonics are all defined relative to the origin $\cO_{\b}$.  In
particular, $r'_{\b}$ and $\hat\bfx'_{\b}$ are the radial and angular
coordinates relative to $\cO_{\b}$ corresponding to a point
$\bfx'$ on the surface $\Sigma_{\b}$.  The integrals over
$\Sigma_{\b}$ define the \emph{multipole moments} of the source
$\varrho_{\b}$, which will be our final quantum variables,
\begin{equation}
\label{multipoles}
Q_{\b,lm}\equiv
\int_{\Sigma_{\b}}d\bfx_{\b}j_{l}(k r_{\b})
Y^{*}_{lm}(\hat\bfx_{\b})\varrho_{\b}(\bfx_{\b}),
\end{equation}
so that
\begin{equation}
\label{phib3}
\phi_{\b}(\bfx_\beta)= {ik}\sum_{lm}Q_{\b,lm} h^{(1)}_{l}(k r_{\b})
Y_{lm}(\hat\bfx_{\b}).
\end{equation}

The field $\phi_{\b}$ viewed from the surface $\Sigma_{\a}$ is a
superposition of solutions to the Helmholtz equation that are regular
at the origin $\cO_{\a}$.  Using translation formulas,
summarized in Ref.~\cite{ref:translation}, the
field generated by object $\Sigma_\beta$ can be written
as function of the coordinate $\bfx_\alpha$, measured from the
origin $\cO_\alpha$, as
\begin{equation}
\label{phib4}
\phi_{\b}(\bfx_\alpha)= {ik}\sum_{lm}Q_{\b,lm} \sum_{l'm'}
\cU^{\alpha\beta}_{l'm'lm}j_{l'}(kr_{\a})Y_{l'm'}(\hat\bfx_{\a}) \, .
\end{equation}
The matrices $\cU^{\alpha\beta}_{l'm'lm}$ are shape and boundary
condition independent and represent the interaction between the
multipoles.  This result, in turn, can be substituted into the
contribution $\widetilde S_{\beta\alpha}$ to the action, leading to
the simple result
\begin{equation}
\label{offdiag}
\widetilde S_{\b\a}[Q_{\a},Q_{\b}]=
{\frac{ik}{2}}\sum_{lml'm'}Q^{*}_{\a,l'm'}\cU^{\alpha\beta}_{l'm'lm}
Q_{\b,lm}+\cc \, .
\end{equation}
Note that the contributions to the action that couple fields and
sources on different objects make no reference to the particular
boundary conditions that characterize the Casimir problem.  They
depend only on the multipole moments of the fields and on the geometry
through the translation matrix $\mathbb{U}^{\a\b}$.

\subsubsection{Self-interaction terms}

We turn to the terms in $\widetilde S_\text{cl}$ where the field
and the source both refer to the same surface, $\Sigma_{\a}$:
\begin{equation}
\label{self1}
\widetilde S_{\a}[\varrho_{\a}]={\frac{1}{2}}
\int_{\Sigma_\a}  d\bfx \left( \varrho^{*}_{\a}(\bfx )
\phi_{\a}(\bfx) + \cc \right) \, .
\end{equation}
For the self-interactions terms, we only use the coordinate
system with origin $\cO_{\a}$ inside the surface
$\Sigma_\a$, and hence drop the label $\alpha$ on the coordinates
in this section.  Since $\phi_\alpha(\bfx)$ is continuous across the
surface, we can regard the $\phi_{\a}$ in Eq.~\eqref{self1} as the field
\emph{inside} $\Sigma_\a$, $\phi_{{\rm in},\a}$, which is a solution
to Helmholtz's equation that must be regular at the origin $\cO_{\a}$,
\begin{equation}
\label{self2}
\phi_{{\rm
in},\a}(\bfx)=\sum_{lm}\phi_{\a,lm}j_{l}(kr)Y_{lm}(\hat\bfx) 
\,  .
\end{equation}
Substituting this expansion into
Eq.~\eqref{self1}, we obtain
\begin{equation}
\label{self3}
\widetilde S_{\a}[\varrho_{\a}]=\frac{1}{2}
\sum_{lm}\left( \phi_{\a,lm}Q^{*}_{\a,lm}+\cc \right)\, ,
\end{equation}
where {the} $Q_{\a,lm}$ are the multipole moments of
the source{s,} defined in the previous subsection.

Finally we relate $\phi_{\a,lm}$ back to the multipole moments of the
source to get an action entirely in terms of the 
$Q_{\a{,lm}}$.  The field
$\phi_{\a,\rm{out}}$ at points \emph{outside} of $\Sigma_{\a}$ obeys
Helmholtz's equation and must equal $\phi_{\a,{\rm in}}$ on the
surface $S$.  Therefore it can be written as $\phi_{\a,{\rm in}}$
\emph{plus a superposition of the regular solutions to the Helmholtz
equation that vanish on $\Sigma_{\a}$}, 
\begin{eqnarray}
\label{outsidereg}
\hspace*{-1.2cm}
\phi_{\a,{\rm out}}(\bfx) &=&\phi_{\a,{\rm in}}(\bfx)+\Delta\phi_{\a}(\bfx)\\
&=&
\phi_{\a,{\rm in}}(\bfx) +\sum_{lm}\chi_{\a,lm}\left(j_{l}(k r)
Y_{lm}(\hat\bfx)
+\sum_{l'm'}{\cal T}^\a_{l'm'l m }(k)h_{l'}^{(1)}(k r)
Y_{l'm'}(\hat\bfx)\right) \, . \nonumber
\end{eqnarray} 
The second term, $\Delta\phi_{\a}$, vanishes on $\Sigma_{\a}$ because
${\mathbb T}^{\a}$ is the scattering amplitude for the Dirichlet
problem.

The field we seek is generated in response to the sources and
therefore falls exponentially (for $k$ {with positive imaginary
  part}) as $r\to\infty$.  Therefore the terms in
Eq.~\eqref{outsidereg} that are proportional to $j_{l}(kr)$ must
cancel.  Comparing Eq.~\eqref{outsidereg} with Eq.~\eqref{self2}, we
conclude that $\chi_{\a,lm}=-\phi_{\a,lm}$, and therefore
\begin{equation}
\label{phiout}
\phi_{\a,{\rm out}}(\bfx) =-\sum_{lm}\phi_{\a,lm}\sum_{l'm'}
{\cal T}^{\a}_{l'm'l m }(k )h_{l'}^{(1)}(k r)Y_{l'm'}(\hat\bfx) \, .
\end{equation}
On the other hand, $\phi_{\a,{\rm out}}(\bfx)$ can be expressed as an
integral over the source as in Eq.~\eqref{phib},
\begin{equation}
\phi_{\a,{\rm out}}(\bfx)=\int_{\Sigma_{\a}} d\bfx' \cG_{0}(\bfx,\bfx'{,k})
\varrho_\alpha(\bfx')  \, .
\end{equation}
Using the partial wave expansion for the free Green's function,
Eq.~\eqref{gf3}, we find
\begin{equation}
\label{outagain}
\phi_{\a, {\rm out}}(\bfx)=ik\sum_{l'm'}Q_{\a,l'm'}h_{l'}^{(1)}(kr)Y_{l'm'}(\hat\bfx) ,
\end{equation}
and comparing with Eq.~\eqref{phiout}, we see that
\begin{equation}
ikQ_{\a,l'm'}=-\sum_{lm}\cT^{\a}_{l'm'lm}(k)\phi_{\a,lm}  ,\nonumber
\end{equation}
or 
\begin{equation}
\phi_{\a,lm}=-ik\sum_{l'm'}[\cT^{\a}]^{-1}_{lml'm'}Q_{\a,l'm'},
\end{equation}
where {$[\mathbb{T}^\alpha]^{-1}$} is the inverse of the Dirichlet
transition matrix $\mathbb{T}^\alpha$.  When this is combined with
Eq.~\eqref{self3}, we obtain the desired expression for the
self-interaction contribution to the action,
\begin{equation}
\label{selfaction}
\widetilde S_{\a}[Q_{\a}]={-\frac{ik}{2}}
\sum_{lml'm'}Q^{ *}_{\a,lm}[\cT^{\a}]^{-1}_{lml'm'}
Q_{\a,l'm'} + \cc \, . 
\end{equation}

\subsection{Evaluation of the Integral over Sources}

Combining Eq.~\eqref{selfaction} with Eq.~\eqref{offdiag}, we obtain
an expression for the action that is a quadratic functional of the
multipole moments of the sources on the surfaces.  The functional
integral Eq.~\eqref{jintegral} can be evaluated by changing variables
from the sources, $\{\varrho_{\a}\}$ to the multipole moments.  The functional
determinant that results from this change of variables can be
discarded because it is a common factor which cancels between
$\mathfrak{Z}_{\cC}$ and $\mathfrak{Z}_{\infty}$.  To
compute the functional integral we analytically continue to imaginary
frequency, $k=i\kappa$, $\kappa>0$,
\begin{equation}
\label{qintegral}
\mathfrak{Z}_{\cC}({i\kappa})=
\prod_{\a=1}^{N}\int \left[{\cD}Q_{\a}\cD Q^{*}_{\a}\right]
\exp\left\{{-\frac{\kappa}{2}}
{\frac{T}{\hbar}}
\sum_{\a}Q_{\a}^{*}[{\mathbb T}^{\a}]^{-1}Q_{\a}
+{\frac{\kappa}{2}}
{\frac{T}{\hbar}}
\sum_{\a\ne\b}Q_{\a}^{*}{\mathbb
U}^{\alpha\beta}Q_{\b}+ \cc \right\} \, ,
\end{equation}
where we have suppressed the partial wave indices.  The functional
integral Eq.~\eqref{qintegral} yields the inverse determinant of a matrix
$\mathbb{M}_{\cC}^{{\alpha\beta}}$ that is composed of the
inverse transition matrices $[{\mathbb T}^\alpha]^{-1}$ on its
diagonal and the translation matrices ${\mathbb U}^{\alpha\beta}$ on the
off-diagonals:
\begin{equation}
\mathbb{M}_{\cC}^{{\alpha\beta}} =
[\mathbb{T^{\alpha}}]^{-1}{\delta_{\alpha\beta}}
- \mathbb{U}^{{\alpha\beta}} {(1-\delta_{\alpha\beta})}\, .
\end{equation}
Finally we substitute into
Eq.~\eqref{casimir2} to obtain the Casimir energy,
\begin{equation}
\label{finalform}
\cE[\cC]=\frac{\hbar c}{\pi}\int_0^\infty d\kappa 
\ln\frac{\det\mathbb{M}_{\cC}(i\kappa)}{\det\mathbb{M}_{\infty}(i\kappa)} \, ,
\end{equation}
where the determinant is taken with respect to the partial wave
indices and the object indices $\alpha$, $\beta$, and
$\mathbb{M}_{\infty}^{\alpha\beta}=[\mathbb{T}^\alpha]^{-1}
\delta_{\alpha\beta}$
is the result of removing the objects to infinite separation, where
the interaction effects vanish.

In the special case of two interacting objects Eq.~\eqref{finalform}
simplifies to
\begin{equation}
\label{finalform2}
\cE_{2}[\cC]=\frac{\hbar c}{\pi}\int_0^\infty d\kappa 
\ln\det(1-{\mathbb T}^{1}\mathbb{U}^{12}\mathbb{T}^{2}\mathbb{U}^{21}) \, ,
\end{equation}
where ${\mathbb T}^{\a}, \a=1,2$, and $\mathbb{U}^{\a\b}$ are the
transition and translation matrices for the two objects. 

\section{Casimir energy from functional integral: Electromagnetic field}

In this section we provide a brief description of the generalization
of the concepts from the previous section to electromagnetic
fields \cite{Emig:2007cf}. Here we start directly with a formulation
in terms of sources, the current and charge densities $\bJ$, $\varrho$.
Using that the EM gauge and scalar potential $[A(\bfx,t) ,\Phi(\bfx,t)]$
are given by 
\begin{equation}
  \label{eq:A-phi-solutions}
  [\bA(\bfx),\Phi(\bfx)]=\int d\bfx' \, G_0(\bfx,\bfx') [\bJ(\bfx'),\varrho(\bfx')]~,
\end{equation}
where now $G_0(\bfx,\bfx')=e^{ik|\bfx-\bfx'|}/(4\pi|\bfx-\bfx'|)$, the
action $S[\bJ]=\int (dk/4\pi) (S_k[\bJ]+S^*_k[\bJ])$ for the currents
densities, defined inside the objects, can be written as
\begin{equation}
  \label{eq:action-currents}
  S_k[\{\bJ_\alpha\}]= \frac{1}{2}\int d \bfx\,  d \bfx' \, 
\sum_{\alpha\beta}\bJ_\alpha^*(\bfx) \, \cG_0(\bfx,\bfx') \, \bJ_\beta(\bfx')\, ,
\end{equation}
where $\cG_0(\bfx,\bfx')= G_0(\bfx,\bfx') - \frac{1}{k^2} \nabla
\otimes \nabla' G_0(\bfx,\bfx')$ is the tensor Green's function. This
is the analogous expression to the one for a scalar field in
Eq.~\eqref{jintegral} with the solution of Eq.~\eqref{phib}
substituted. Next we must constrain the currents to be {\it induced}
sources that depend on shape and material of the objects. Formally
this is achieved by integrating over currents, inserting constraints
to ensure that the currents in vacuum simulate the correct induction
of microscopic polarization $\bP_\alpha$ and magnetization
$\bM_\alpha$ (from all multipoles) inside the dielectric objects in
response to an incident wave.

Let us consider one object.  First, the induced current is
$\bJ_\alpha=-ik\bP_\alpha+\nabla\times \bM_\alpha$, and since
$\bP_\alpha=(\epsilon_\alpha-1)\bE$, $\bM_\alpha=(1-1/\mu_\alpha)\bB$,
it can be expressed in terms of the total fields $\bE$, $\bB$ inside
the object as
\begin{equation}
  \label{eq:J-condition}
  \bJ_\alpha = -ik (\epsilon_\alpha-1) \bE + \nabla \times 
[(1-1/\mu_\alpha) \bB] \, .
\end{equation}
Second, the total field inside the object must consist
of the field generated by $\bJ_\alpha$ and the incident field
$\bE_0(\{\bJ_\alpha, \bbS^\alpha\},\bfx)$ that has to impinge on the object to
induce $\bJ_\alpha$, so that
\begin{equation}
  \label{eq:total_E}
  \bE(\bfx)= \bE_0(\{\bJ_\alpha, \bbS^\alpha\},\bfx) +
 ik \int d \bfx' \, \cG_0(\bfx,\bfx') \, \bJ_\alpha(\bfx') \, .
\end{equation}
The incident field depends on the current density to be induced and
on the scattering matrix $\bbS^\alpha$ of the object, which connects
the incident wave to the scattered wave. It is fully
specified by the multipole moments of $\bJ_\alpha$ (see below for
details).  Substituting Eq.~\eqref{eq:total_E} and $\bB=(1/ik)
\nabla\times \bE$ into Eq.~\eqref{eq:J-condition} yields
a self-consistency condition that constrains the current
$\bJ_\alpha$. If one writes this condition as
$\cC_\alpha[\bJ_\alpha]=0$ for each object, the functional
integration over the currents constrained this way for all objects
yields the partition function
\begin{equation}
  \label{eq:partition-fct}
  \cZ=\int \prod_\alpha \cD \bJ_\alpha \prod_{\bfx\in V_\alpha} 
\delta(\cC_\alpha[\bJ_\alpha(\bfx)])
\exp\left(i S[\{\bJ_\alpha\}]\right) \, .
\end{equation}

It is instructive to look at  two compact objects at a
distance $d$, measured between the (arbitrary) origins ${\cal
  O}_\alpha$ inside the objects.  In this case the action of
Eq.~\eqref{eq:action-currents} is
\begin{eqnarray}
  \label{eq:S-local}
  S_k[\{\bJ_\alpha\}] &=& \frac{1}{2} \sum_{\alpha\neq\beta}
\int d\bfx_\alpha \, \bJ_\alpha^*(\bfx_\alpha) \frac{1}{ik} 
\bE_\beta(\bfx_\alpha-d_\alpha\hz)\\
&+& \frac{1}{2} \sum_\alpha \int \! d\bfx_\alpha  d\bfx'_\alpha  \,
\bJ_\alpha^*(\bfx_\alpha) \, \cG_0(\bfx_\alpha,\bfx'_\alpha)\, \bJ_\alpha(\bfx'_\alpha)
\nonumber \, ,
\end{eqnarray}
where we have substituted the electric field $\bE_\alpha(\bfx_\alpha)=
ik \int d \bfx'_\alpha \, \cG_0(\bfx_\alpha,\bfx'_\alpha) \,
\bJ_\alpha(\bfx'_\alpha) $ and the fields are measured now in local
coordinates so that $\bfx={\cal O}_\alpha+\bfx_\alpha$, and
$d_\alpha=d$ ($-d$) for $\alpha=1$($2$).  The off-diagonal terms in
Eq.~\eqref{eq:S-local} represent the interaction between the currents
on the two materials.  A natural way to decompose the interaction between 
charges is to use the multipole expansion. For
each body we define magnetic and electric multipoles as
\begin{eqnarray}
  \label{eq:mulitpole-moments}
  Q^\alpha_{\textsc{m},lm}&\!=\!& \frac{k}{\lambda} \!\int\!\! d \bfx_\alpha \, 
\bJ_\alpha(\bfx_{\alpha})
\nabla\times[\bfx_\alpha j_l(kr_\alpha) Y^*_{lm}(\hr_\alpha)]  \\
   Q^\alpha_{\textsc{E},lm}&\!=\!& \frac{1}{\lambda}\!\int\!\! d \bfx_\alpha \, 
\bJ_\alpha(\bfx_{\alpha})
\nabla\!\times\!\nabla\times[\bfx_\alpha j_l(kr_\alpha) Y^*_{lm}(\hr_\alpha)] \nonumber\, ,
\end{eqnarray}
for $l\ge 1$, $|m|\le l$, where $\lambda=\sqrt{l(l+1)}$, $j_l$ are
spherical Bessel functions and $Y_{lm}$ spherical harmonics.  We
change variables from currents to multipoles in the functional
integral and, as the final step in our quantization, integrate over all
multipole fluctuations on the two objects weighted by the effective
action,
\begin{equation}
  \label{eq:S_eff_mp}
S^\text{eff}_k[\{Q^\alpha_{lm}\}] =\frac{1}{2} \frac{i}{k} \sum_{lml'm'} \big\{
Q^{1*}_{lm} \, U^{12}_{lml'm'} \, Q^2_{l'm'}  
+  Q^{2*}_{lm} \, U^{21}_{lml'm'} \, Q^1_{l'm'} 
+ \!\!\sum_{\alpha=1,2} Q^{\alpha *}_{lm} \, [-T^\alpha]^{-1}_{lml'm'} \, Q^\alpha_{l'm'} 
\big\} \, ,
\end{equation}
with
$Q^\alpha_{lm}=(Q^\alpha_{\textsc{M},lm},Q^\alpha_{\textsc{E},lm})$.
Formally, this action resembles that for a scalar field. The vector
nature of the EM field is reflected by the presence of two different
polarizations and corresponding electric and magnetic multipoles.  Let
us discuss the terms appearing in Eq.~\eqref{eq:S_eff_mp} and sketch
its derivation.

{\it Off-diagonal terms} --- We need to know the electric fields in
Eq.~\eqref{eq:S-local} exterior to the source that generates them.
They can be represented in terms of the multipoles as
$\bE_\beta(\bfx_\beta)=-k \sum_{lm} Q^\beta_{lm}
\bpsi^\text{out}_{lm}(\bfx_{\beta})$ where
$\bpsi^\text{out}_{lm}(\bfx_{\beta})$ are \emph{outgoing} vector
solutions of the Helmholtz equation in the coordinates of object
$\beta$ \footnote{The two components 
of  $\bpsi^\text{reg}_{lm}$ are given by $1/k$ times the weights for
  $\textsc{E}$- and $\textsc{M}$-multipoles of
  Eq.~\eqref{eq:mulitpole-moments} with $Y^*_{lm}$ replaced by
  $Y_{lm}$. Similarly, $\bpsi^\text{out}_{lm}$ have the same
  expressions upon substituting Bessel by Hankel functions, $j_l \to
  h_l^{(1)}$.}.  We would like to express the currents
$\bJ_\alpha^*$ in Eq.~\eqref{eq:S-local} also in terms of multipoles.
The difficulty in doing so is that the electric field is expressed in
terms of {\it outgoing} partial waves in the coordinates of object
$\beta$, while according to Eq.~\eqref{eq:mulitpole-moments}, the
multipoles involve partial waves $\bpsi^\text{reg}_{lm}(\bfx_{\alpha})$
that are \emph{regular} at the origin ${\cal O}_{\alpha}$, in the
coordinates of object $\alpha$.  Going from the
outgoing to the regular vector solutions and changing the coordinate
system involves a translation and change of basis which can be
expressed as $\bpsi^\text{out}_{lm}(\bfx_\alpha\pm
d\hz)=\sum_{l'm'}U^\pm_{l'm'lm} \bpsi^\text{reg}_{l'm'}(\bfx_\alpha)$
where the {\it universal} (shape and material independent) translation matrices
$\bbU^{21}$ and $\bbU^{12}$ represent the interaction between the
multipoles. For fixed $(lm)$, $(l'm')$, they are $2\times 2$ matrices
(magnetic and electric multipoles), and functions of $kd$ only. Their
explicit form is known but not provided here to save space
\cite{wittmann}; they fall off with $kd$ according to classical
expectations for the EM field. Then the electric field becomes
$\frac{1}{ik}\bE_\beta(\bfx_\alpha\pm d\hz)= \sum_{lm} \phi^\beta_{lm}
\bpsi^\text{reg}_{lm}(\bfx_{\alpha})$ with $\phi^\beta_{lm}=i\sum_{lm}
U^{21(12)}_{lml'm'} Q^\beta_{l'm'}$, and the integration in
Eq.~\eqref{eq:S-local} leads, using Eq.~\eqref{eq:mulitpole-moments},
to the off-diagonal terms in Eq.~\eqref{eq:S_eff_mp}.

{\it Diagonal terms} --- The self-action, given by the second term of
Eq.~\eqref{eq:S-local}, is more interesting and more challenging.  It
can be expressed in terms of multipoles if we use the constraint for
the currents, Eqs.~\eqref{eq:J-condition} and \eqref{eq:total_E}. To
do so, we first note that in scattering theory one usually knows the
incident solution and would like to find the outgoing scattered
solution.  They are related by the $S$-matrix.  Here the
situation is slightly different.  We seek to relate a regular solution
$\bE_0(\bfx_{\alpha})=ik\sum_{lm}\phi_{0,lm}\bpsi^\text{reg}_{lm}(\bfx_\alpha)$
and the outgoing scattered solution,
$\bE_\alpha(\bfx_{\alpha})=-k\sum_{lm}Q^{\alpha}_{lm}
\bpsi^\text{out}_{lm}(\bfx_\alpha)$, generated by the currents in the
material --- a relation determined by the T-matrix,
$\bbT^\alpha\equiv (\bbS^\alpha-\bbI)/2$ --- schematically
$i \bQ^\alpha= \bbT^\alpha\bphi_0$ \footnote{Our relation between S- and T-matrix follows
  Ref.~\cite{Waterman:1971a} and hence deviates from usual conventions
  by a factor $i$.} \cite{Waterman:1971a}.  We
face the inverse problem of determining $\phi_{0,lm}$ for known
scattering data $Q^\alpha_{lm}$, hence,
\begin{equation}
  \label{eq:incident_amplitudes}
    \phi_{0,lm}= i \sum_{l'm'} \, [T^\alpha]^{-1}_{lml'm'} Q^\alpha_{l'm'} \,
\end{equation}
so that the incident field is given in terms of the S-matrix, as
indicated in Eq.~\eqref{eq:total_E}. Next, we express the self-action
of the currents inside a body (the second term of
Eq.~\eqref{eq:S-local}), as $S^{\alpha}_k[\bJ_{\alpha}] = \frac{1}{2}
\int d \bfx_\alpha [ \bE \bD^* - \bB \bH^* - ( \bE_0 \bD_0^* - \bB_0
\bH_0^* ) ]$, the change of the field action that results from placing
the body into the {\it fixed} (regular) incident field $\bE_0=\bD_0$,
$\bH_0=\bB_0$, where $\bE$, $\bH$ and $\bD$, $\bB$ are the new total
fields and fluxes in the presence of the body. Using
$\bD=\epsilon_\alpha \bE$, $\bH=\mu_\alpha^{-1}\bB$ inside the body
and Eq.~\eqref{eq:J-condition}, straightforward manipulations lead to
the simple self-action $S^{\alpha}_k[\bJ_{\alpha}]=-\frac{1}{2ik} \int
d \bfx_\alpha \bJ^*_\alpha \bE_0(\{\bJ_\alpha,\bbS^\alpha\})$.  If we
substitute the regular wave expansion for $\bE_0$ with coefficients of
Eq.~\eqref{eq:incident_amplitudes} and integrate by using
Eq.~\eqref{eq:mulitpole-moments}, we get Eq.~\eqref{eq:S_eff_mp}.

Having established the action for electric and magnetic multipoles,
Eq.~\eqref{eq:mulitpole-moments}, the Casimir energy follows in
complete analogy to the scalar case by integrating over the
multipoles.  Hence, the Casimir energy for two objects is again given
by Eq.~\eqref{finalform2} with an additional factor of $1/2$ since
the electromagnetic field is real valued.

\section{Applications: Scalar field}
\label{appl}

In this section we give a few typical applications of our method for a
scalar field.  We consider a {\it real} scalar field fluctuating in
the space between two spheres on which Robin boundary conditions,
$\phi - \lambda_\alpha \partial_n \phi=0$, are imposed.  Because a
real field has half the oscillation modes of a complex field, the
Casimir energy in Eq.~(\ref{finalform2}) must be divided by 2, giving
\begin{equation}
\label{finalform2real}
\cE_{2}[\cC]=\frac{\hbar c}{2\pi}\int_0^\infty d\kappa 
\ln\det(1-{\mathbb T}^{1}\mathbb{U}^{12}\mathbb{T}^{2}\mathbb{U}^{21}) \, .
\end{equation}
We allow for different Robin parameters $\lambda_{1,2}$ at the spheres
of radius $R$.  This choice allows us to study Dirichlet
($\lambda/R\to 0$) and Neumann ($\lambda/R\to\infty$) boundary
conditions on separate spheres as special cases. We obtain the Casimir
energy as a series in $R/d$  for large separations
$d$ and numerically at all separations.  A comparison of the two
approaches allows us to measure the rate of convergence of our
results.  We find that for Robin boundary conditions the sign of the
force depends on the ratios $\lambda_{\a}/R$ and on the
separation $d$.

\subsection{Interaction of two spheres with Robin boundary conditions:  
general considerations}

The Robin boundary condition $\phi - \lambda_\alpha \partial_n \phi=0$
allows a continuous interpolation between Dirichlet and Neumann
boundary conditions. Since the radius of the sphere introduces a
natural length scale, it is convenient to replace $\lambda$
by a dimensionless variable, $\zeta{_{\a}}\equiv
\lambda{_{\a}}/R$.  For
$\zeta_\alpha>0$, the modulus of the field is suppressed if the
surface is approached from the outside, while for $\zeta_\alpha<0$ it
is enhanced. Hence, for negative $\zeta_\alpha$ bound surface states
can be expected.  All the information about the shape of the
object and the boundary conditions at its surface is provided by the
$\mathbb{T}$-matrix.  For spherically symmetric objects the
$\mathbb{T}$-matrix is diagonal and is completely specified by phase
shifts $\delta_{l}{(k)}$ that do not depend on $m$,
\begin{equation}
  \label{eq:spheres-t-from-shifts}
  \cT_{lml'm'}(k)=\delta_{ll'}\delta_{mm'} \frac{1}{2} 
\left( e^{2i\delta_l(k)}-1\right) \, . 
\end{equation}
In the discussion of the $\mathbb{T}$-matrix for an individual
object we {again} suppress the label $\alpha$.  The phase shifts
for Robin boundary conditions are
\begin{equation}
  \label{eq:sphere-phases}
  \cot \delta_l(k) = \frac{n_l(\xi)-\zeta\xi\, n'_l(\xi)}
{j_l(\xi)-\zeta\xi\, j'_l(\xi)} \, ,
\end{equation}
where $\xi=kR$ and $j_l~(n_l)$ are spherical Bessel functions of
first(second) kind. To apply Eq.~\eqref{finalform2real}, we have to
evaluate the matrix elements of the transition matrices for imaginary
frequencies $k=i\kappa$.  Using  $j_l(iz)=i^l
\sqrt{\pi/(2z)} I_{l+1/2}(z)$ and $h^{(1)}_l(iz)= -i^{-l} \sqrt{2/(\pi
  z)} K_{l+1/2}(z)$, we obtain for the $\mathbb{T}$-matrix elements
\begin{equation}
  \label{eq:spheres-t-matrix-imag}
  \cT_{lmlm}(i\kappa)=(-1)^l \frac{\pi}{2} \frac{(1/\zeta+1/2) I_{l+1/2}(z)
-z I'_{l+1/2}(z)}{(1/\zeta+1/2) K_{l+1/2}(z)
-z K'_{l+1/2}(z)} \, ,
\end{equation}
where $z\equiv \kappa R$.

For two spherical objects we can assume that the center-to-center
distance vector is parallel to the $z$-axis. Then the translation
matrices simplify. For imaginary frequencies the translation matrix
elements become
\begin{eqnarray}
  \label{eq:spheres-transl-matrix-imag}
\cU^{\left\{\begin{matrix}12\\ 21\end{matrix}\right\}}_{l'mlm}(d)&=& 
- (-1)^{m} i^{-l'+l} \sqrt{(2l+1)(2l'+1)}\\
&\times& \sum_{l''} \,
 (\pm 1)^{l''} (2l''+1)
\begin{pmatrix}l&l'&l''\\0&0&0\end{pmatrix}
\begin{pmatrix}l&l'&l''\\m&-m&0\end{pmatrix}
K_{l''+1/2}(\kappa d) \, , \nonumber
\end{eqnarray}
where $d$ is the separation distance.

An analysis of the $\mathbb{T}$-matrix shows that it
has poles for  $-1 < \zeta <0$. For any
$\zeta$ in this interval, there exists a finite number of bound
states, which increases as $\zeta \to 0$. In the following, we
restrict to $\zeta\ge 0$ and leave the study of interactions in the
presence of bound states to a future publication.

The special case of spheres with Dirichlet boundary conditions has
been studied in Ref.~\cite{Bulgac:2005ku}. For two spheres of equal
radius, the matrix $\sum_{l''}A_{ll''}^{(m)}A_{l''l'}^{(m)}$ in the
notation of Ref.~\cite{Bulgac:2005ku} is proportional to our
$\mathbb{T}^{1}\mathbb{U}^{12}\mathbb{T}^{2}\mathbb{U}^{21}$ times
$K_{l+1/2}(\kappa R)/K_{l'+1/2}(\kappa R)$. It is easy to see that
this proportionality factor drops out in the final result for the
energy if one uses $\ln \det = \tr \ln$ in Eq.~\eqref{finalform2real}
and expands the logarithm around unity. Thus we agree with the results
given in Ref.~\cite{Bulgac:2005ku}.

\subsection{Asymptotic expansion for large separation}

In this section we consider the Casimir interaction between two
spheres due to a scalar field obeying Robin boundary conditions,
allowing for a different parameter $\lambda_{1,2}$ on each sphere.
The Casimir energy can be developed in an asymptotic expansion in
$R/d$ using $\ln \det = \tr \ln$ in Eq.~\eqref{finalform2real}.
Expanding the logarithm in powers of $\mathbb{N}= \mathbb{T}^{1}
\mathbb{U}^{12} \mathbb{T}^{2} \mathbb{U}^{21}$, since the
$\mathbb{T}$-matrix has no poles in the region of interest
we get
\begin{equation}
  \label{eq:spheres-energy-real}
  \cE = - \frac{\hbar c}{2\pi} \int_0^\infty d\kappa \sum_{p=1}^\infty 
\frac{1}{p} \tr\left(\mathbb{N}^p\right) \, .
\end{equation}
We have performed the matrix operations using {\tt Mathematica}. The
scaling of the $\mathbb{T}$-matrix at small $\kappa$ shows that the 
$p^\text{th}$ power of $\mathbb{N}$ (corresponding to $2p$
scatterings) become{s} important at order $d^{-(2p+1)}$. Partial waves of
order $l$ start to contribute at order $d^{-(3+2l)}$ if the
$\mathbb{T}$-matrix is diagonal in $l${,} which is the case for
spherically symmetric objects. Hence the leading terms with $p=1$ and
$l=0$ yield the exact energy to order $d^{-4}$. In the following we
will usually restrict the expansion to $p \le 3$, $l \le 2$,
yielding the interaction to order $d^{-8}$.

The large distance expansion of the Casimir energy
can be written as
\begin{equation}
\label{eq:spheres-energy-large-d}
\cE = \frac{\hbar c }{\pi} \frac{1}{d}
\sum_{j=3}^\infty b_j \left(\frac{R}{d}\right)^{j-1} \, ,
\end{equation}
where $b_j$ is the coefficient of the term $\sim d^{-j}$.  These
coefficients can be computed for general Robin boundary conditions
\cite{Emig+07}. Here we restrict to the limiting cases of Dirichlet
and Neumann boundary conditions. An interesting property is that some
coefficients $b_j$ go to zero for $\lambda_\alpha\to\infty$, which
corresponds to Neumann boundary conditions. If both $\lambda_\alpha$
go to infinity, the coefficients $b_j$ vanish for $j=1,\ldots, 6$, so
that the leading term in the Casimir energy is $\sim d^{-7}$ with a
negative amplitude. Hence, Neumann boundary conditions lead to an
attractive Casimir-Polder power law, as is known from electromagnetic
field fluctuations.  This result can be understood from the absence of
low-frequency $s$-waves for Neumann boundary conditions. It is clearly
reflected by the low frequency expansion of the ${\mathbb T}$-matrix,
which has a vanishing amplitude for $\lambda_\alpha\to\infty$ if
$l=0$.  If one $\lambda_\alpha=0$ and the other goes to infinity, only
the coefficients $b_3$ and $b_4$ vanish so that the energy scales as
$d^{-5}$ with a {\it positive} amplitude.  In general, one has $b_3<0$ for
$\lambda_\alpha\ge 0$, so that at asymptotic distances the Casimir force is
attractive for all non-negative finite $\lambda_\alpha$, and
for $\lambda_\alpha$ both infinite.  It is repulsive if one
$\lambda_\alpha$ is finite and the other infinite, i.e., if one sphere
obeys Neumann boundary conditions. However, at smaller distances the
interaction can change sign depending on $\lambda_\alpha$, as shown
below.

More precisely, we obtain the following results. If both
$\lambda_\alpha=0$, the field obeys Dirichlet conditions at the two
spheres and the first six coefficients are
\begin{equation}
\label{eq:coeff_D-D}
b_3=-\frac{1}{4}, \quad b_4=-\frac{1}{4}, \quad b_5=-\frac{77}{48}, \quad 
b_6=-\frac{25}{16}, \quad 
b_7=-\frac{29837}{2880}, \quad b_8=-\frac{6491}{1152} \, .
\end{equation}
If Neumann conditions are imposed on both surfaces, the coefficients are
\begin{equation}
\label{eq:coeff_N-N}
b_3=0, \quad b_4=0, \quad b_5=0, \quad 
b_6=0, \quad b_7=-\frac{161}{96}, \quad b_8=0, \quad b_9=-\frac{3011}{192}
\quad b_{10}=-\frac{175}{128}\, ,
\end{equation}
clearly showing that the asymptotic interaction has a Casimir-Polder
power law $\sim {{\cal O}(}d^{-7}{)}$. Also, as in the
electromagnetic case, the next to leading order ${\cal O}(d^{-8})$
vanishes \cite{Emig:2007cf}.  Therefore we have
included the two next terms of the series.
For mixed Dirichlet/Neumann boundary conditions, we obtain
\begin{equation}
  \label{eq:coeff_D-N}
    b_3=0, \quad b_4=0, \quad b_5=\frac{17}{48}, \quad 
b_6=\frac{11}{32}, \quad b_7=\frac{663}{160}, \quad b_8=\frac{235}{144} \, .
\end{equation}
It is important to note that the series in
Eq.~(\ref{eq:spheres-energy-large-d}) is an asymptotic series and
therefore cannot be used to obtain the interaction at short distances.

\subsection{Numerical results for Robin boundary conditions on two spheres 
at all separations}

The primary application of our analysis is to compute the Casimir
energy and force to high accuracy over a broad range of distances.
However, to obtain the interaction at all distances,
Eq.~\eqref{finalform2real} has to be evaluated numerically.  We shall
see that the domain where our method is \emph{least} accurate is when
the two surfaces approach one another.  That is the regime where
semiclassical methods like the proximity force approximation (PFA)
become exact.  Because of its role in this limit, and because it is
often used (with little justification) over wide ranges of
separations, it is important to compare our calculations with the PFA
predictions.

In the proximity force approximation, the energy is obtained as an
integral over infinitesimal parallel surface elements at their local
distance $L$, measured perpendicular to a surface $\Sigma$
that can be one of the two surfaces of the objects, or an auxiliary
surface placed between the objects. The PFA approximation for the
energy is then given by
\begin{equation}
  \label{eq:energy-pfa}
  \cE_\text{PFA}=\frac{1}{A}\int_\Sigma \cE_\|(L) dS \, ,
\end{equation}
where $\cE_\|(L)/A$ is the energy per area for two parallel plates
with distance $L$. The Casimir energy for parallel plates with Robin
boundary conditions has been obtained as function of
$\lambda_\alpha/L$ in Ref.~\cite{Emig+07}.  The behavior of the PFA at
asymptotically small or large $\lambda_{\a}/L$ determines the Casimir
interaction as $L\to 0$.  For all non-zero values of $\lambda_{1,2}$,
we take $\lambda_{\a}/L \to \infty$, but for the Dirichlet case,
$\lambda=0$, the limit $\lambda_{\a}/L\to 0$ applies. For parallel
plates with Robin boundary conditions, in the limit
$\lambda_{1,2}/L\gg 1$ we obtain the result for Neumann boundary
conditions on both plates,
\begin{equation}
  \label{eq:pp-amp-rep}
  \Phi(\lambda_1/L,\lambda_2/L) \to \Phi_0^- = -\frac{\pi^2}{1440} \, ,
\end{equation} 
and for $\lambda_{1,2}/L\ll 1$ we obtain the identical result for plates
with Dirichlet conditions.  Finally for 
$\lambda_{1,2}/L\ll 1\ll
\lambda_{2,1}/L$, we obtain the parallel plate result for unlike
(Dirichlet/Neumann) boundary conditions,
\begin{equation}
  \label{eq:pp-amp-att}
  \Phi(\lambda_1/L,\lambda_2/L) \to \Phi_0^+ = -\frac{7}{8} \Phi_0^- =
\frac{7\pi^2}{11520} \, .
\end{equation}
The last case is relevant at short distances if one of the
$\lambda_\alpha=0$.  For two spheres of radius $R$ and
center-to-center separation $d$ with Robin boundary conditions, the
PFA results can now be obtained easily from Eq.~(\ref{eq:energy-pfa}). 
In terms of the surface-to-surface distance $L=d-2R$, we get
\begin{equation}
\label{eq:pfa-spheres}
\cE_\text{PFA} =  \Phi_0^\pm \, \frac{\pi}{2}
\frac{R \, \hbar c}{(d-2R)^2} \, ,
\end{equation}
where the $+$ applies if one and only one $\lambda_\alpha=0$, and the
$-$ in all other cases. Hence, at small separation the interaction
becomes independent of $\lambda_\alpha$, in the sense that it only
depends on whether one $\lambda_\alpha$ is zero. 

With the results obtained above, we can analyze the sign of the
interaction between plates and spheres at both asymptotically large
and small distances.  Since the PFA result is expected to hold in the
limit where the distance tends to zero, Eq.~(\ref{eq:pfa-spheres})
predicts the sign of the interaction between spheres in the limit of
vanishing distance.  In the limit of large distances, we can compare
the results for parallel planes from Eqs.~(\ref{eq:pp-amp-rep}) and
(\ref{eq:pp-amp-att}) to our calculations for two spheres.  We find
that the {\it sign} of the asymptotic interaction depends on the
choice for $\lambda_\alpha$ and is {\it identical} for plates and
spheres.  Hence, we obtain a complete characterization of the sign of
the interaction at asymptotically large and small distances for the
plate and sphere geometry, which is summarized in
Table~\ref{tab:sign}. However, as we have seen above, the power law
decay at large distance is quite different for plates and spheres.

\begin{table}[ht]
\begin{center}
\begin{tabular}{|c|c|c|c|l|}
\hline
$\lambda_1$ & $\lambda_2$ & $L \to 0$ & $L \to\infty$ & remark \\  
\hline
0 & 0 & $-$ & $-$ & $-$ for all $L$ \\
$\infty$ & 0 & $+$ & $+$ & $+$ for all $L$ \\
$\infty$ & $\infty$ & $-$ & $-$ & $-$ for all $L$ \\
$]0,\infty[$ & $]0,\infty[$ & $-$ & $-$ & $+$ at intermediate $L$ for \\
& & & & large enough ratio of $\lambda_1$, $\lambda_2$.\\
& & & & (for plates: $\lambda_1/\lambda_2$ or
$\lambda_2/\lambda_1\gtrsim 2.8$)\\
$]0,\infty[$ & 0 & $+$ & $-$ & \\
$]0,\infty[$ & $\infty$ & $-$ & $+$ & \\
\hline
\end{tabular}
\caption{The sign of the Casimir force between two plates and two
spheres with Robin boundary conditions at asymptotically small and large
surface-to-surface distance $L$. The sign in these two limits is
identical for plates and spheres.  Here ``$-$'' and ``$+$'' indicate
attractive and repulsive forces, respectively.}
\label{tab:sign}
\end{center}
\end{table}

\subsubsection{Casimir forces for all separations}

To go beyond the analytic large distance expansion, we compute
numerically the interaction between two spheres of the same radius $R$
with Robin boundary conditions. Guided by the classification of the
Casimir force according to its sign at small, intermediate and large
separations, we discuss the six different cases listed in
Table~\ref{tab:sign}. Our numerical approach starts from
Eq.~\eqref{finalform2real}.  Using the matrix elements
of Eq.~(\ref{eq:spheres-t-matrix-imag}) and
Eq.~(\ref{eq:spheres-transl-matrix-imag}){,} we compute the
determinant and the integral over imaginary frequency numerically.  We
truncate the matrices at a finite multipole order $l$ so that they
have dimension $(1+l)^2 \times (1+l)^2$, yielding a series of
estimates $\cE^{(l)}$ for the Casimir energy.

$\cE^{(1)}$ gives the exact
result for asymptotically large separations, while for decreasing
separations an increasing number of multipoles has to be included.
The exact Casimir energy at all separations is obtained by
extrapolating the series $\{\cE^{(l)}\}$ to $l\to\infty$. We observe an
exponentially fast convergence as $|\cE^{(l)} - \cE|\sim
e^{-\delta(d/R-2)l}$, where $\delta$ is a constant of order
unity. Hence, as the surfaces approach each other for $d\to 2R$, the
rate of convergence tends to zero. However, we find that the
first $l=20$ elements of the series are sufficient to obtain accurate
results for the energy at a separation with $R/d=0.48$, corresponding
to a surface-to-surface distance of the spheres of $L=0.083 R$, i.e.,
approximately $4\%$ of the sphere diameter. In principle our approach
can be extended to even smaller separations by including higher order
multipoles. However, at such small separations semiclassical
approximations like the PFA start to become accurate and can be also
used.

\begin{figure}[ht]
\includegraphics[scale=0.35]{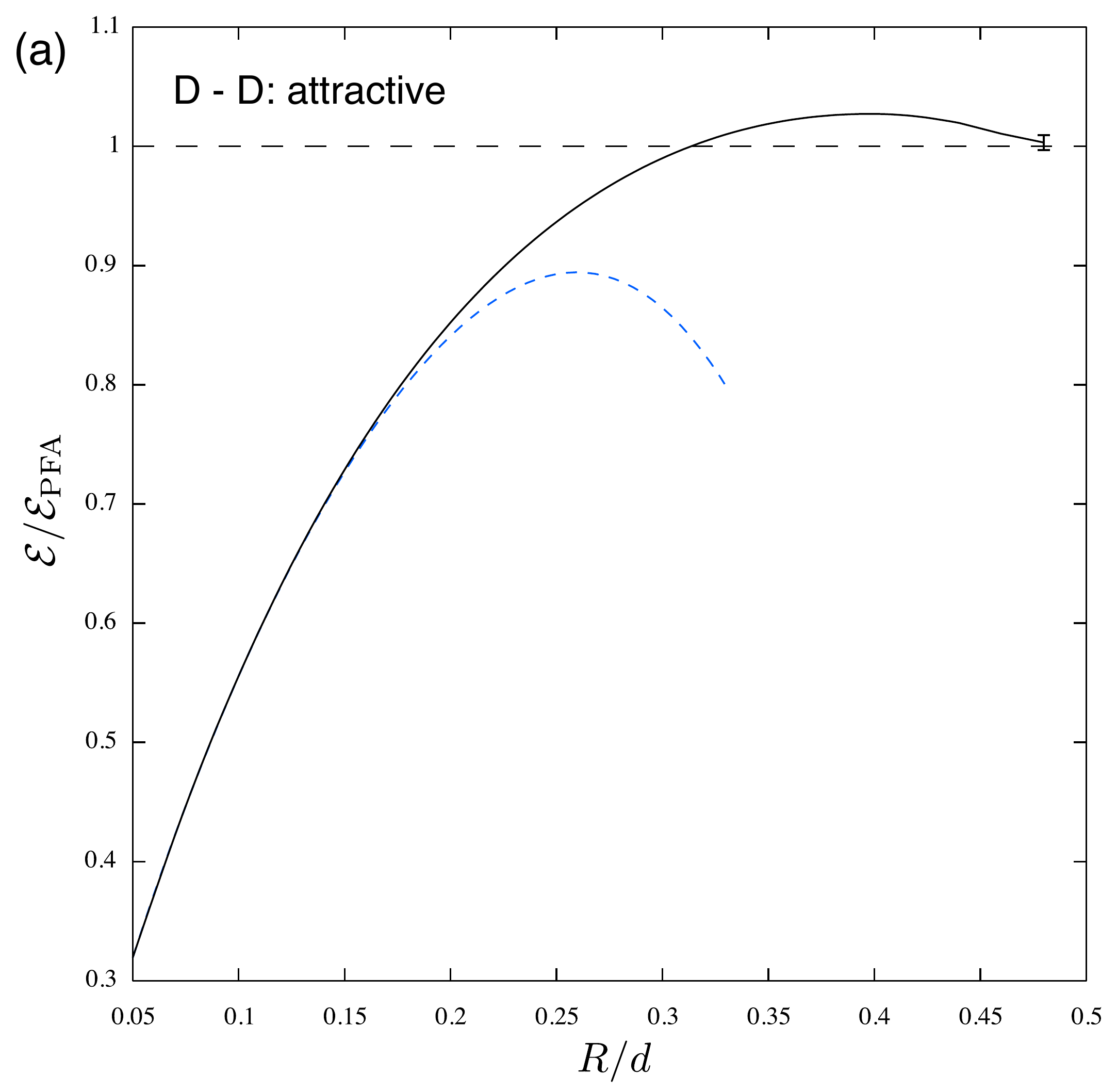}
\includegraphics[scale=0.35]{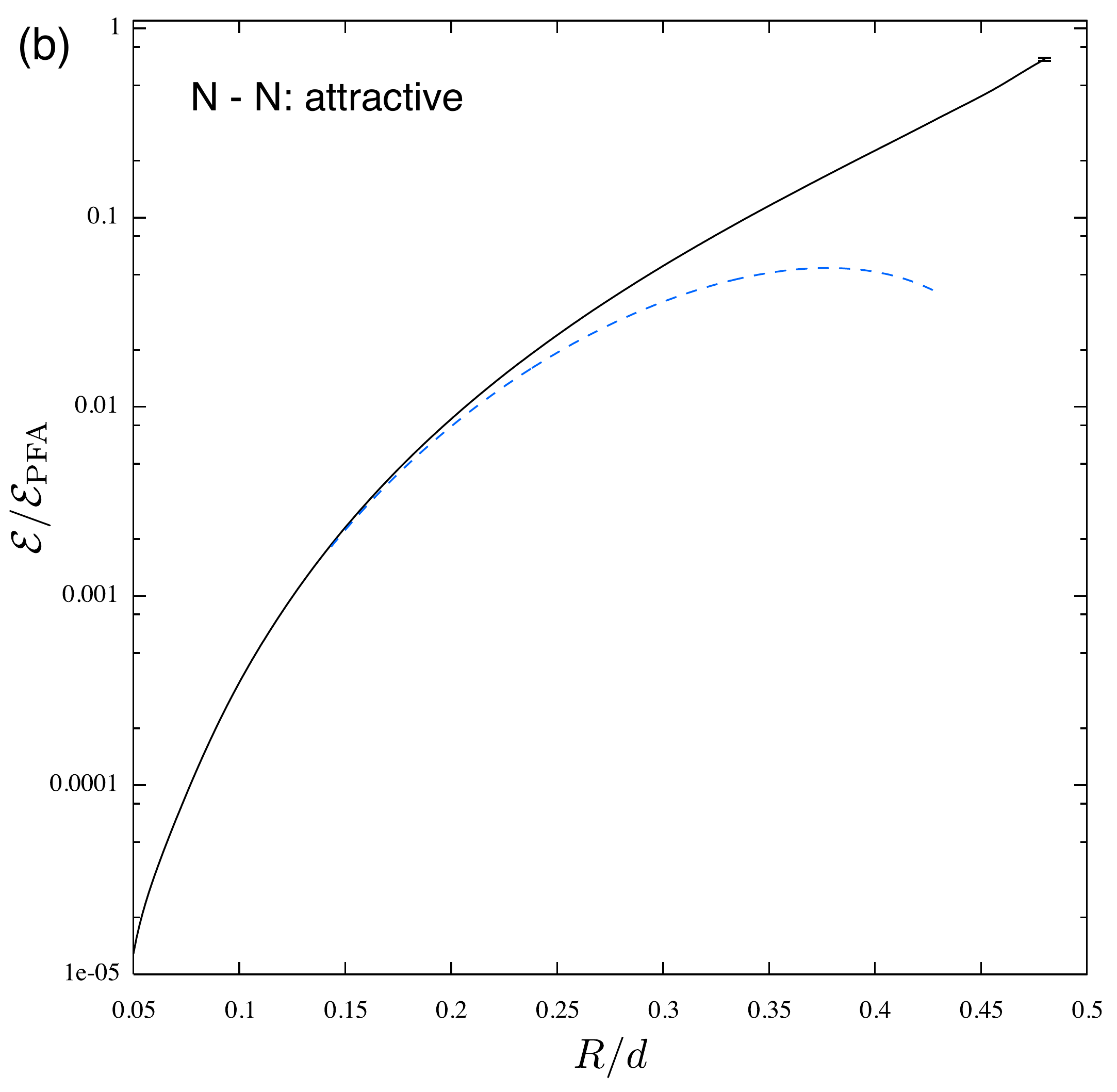}
\includegraphics[scale=0.35]{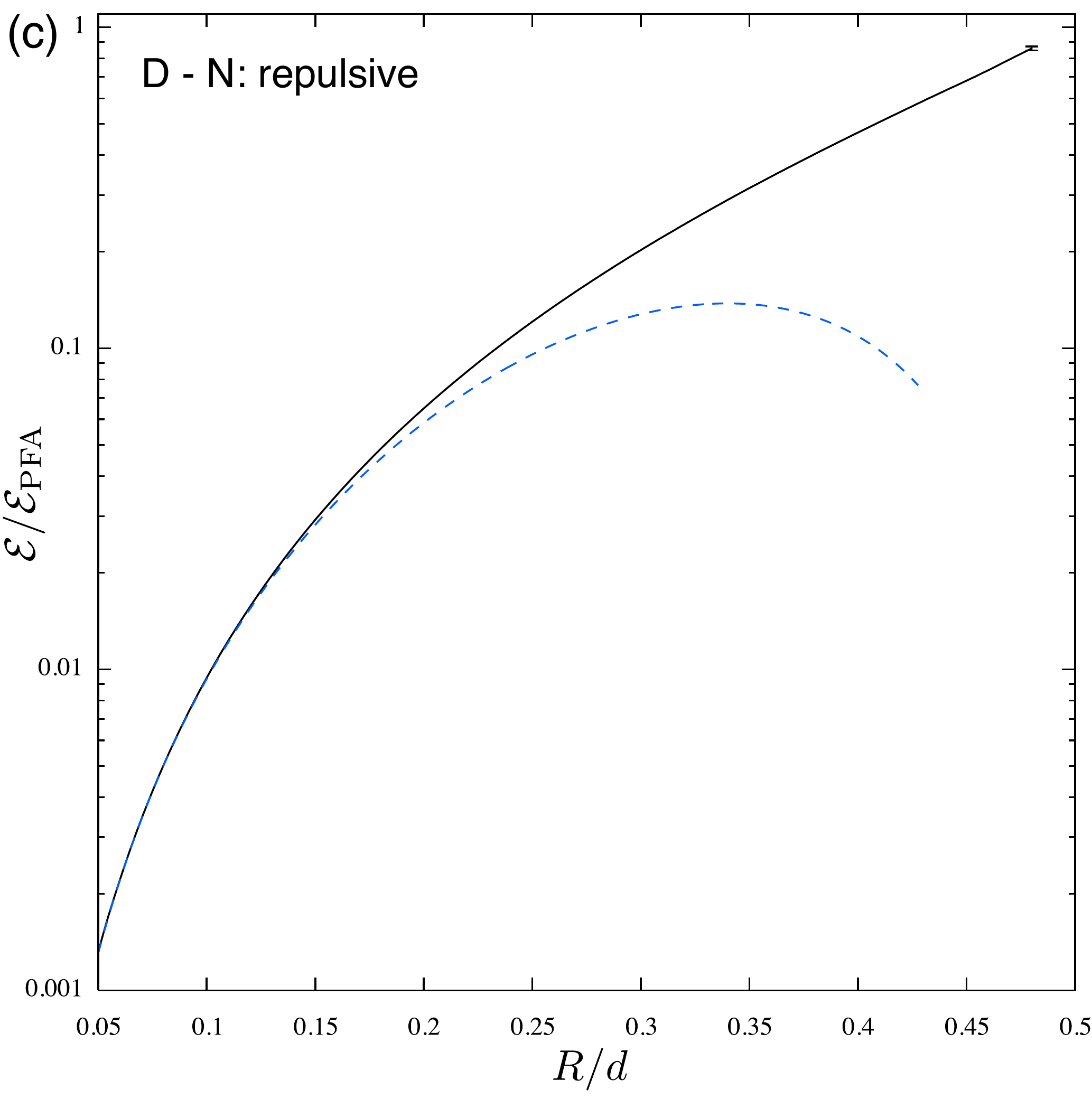}
\hspace*{-2cm}
\parbox[b]{0.62\linewidth}{
\caption{Casimir energy for two spheres of radius $R$ and
  center-to-center distance $d$: (a) Dirichlet boundary conditions for
  both spheres, (b) Neumann boundary conditions for both spheres, (c)
  Spheres with different boundary conditions (one Dirichlet, one
  Neumann).  The energy is scaled by the PFA estimate of
  Eq.~(\ref{eq:pfa-spheres}). The solid curves are obtained by
  extrapolation to $l\to\infty$. For the smallest separation, the
  extrapolation uncertainty is maximal and indicated by an error
  bar. The dashed curves represent the asymptotic large distance
  expansion given in Eq.~(\ref{eq:spheres-energy-large-d}) with the
  coefficients of Eqs.~(\ref{eq:coeff_D-D}), (\ref{eq:coeff_N-N}) and
  (\ref{eq:coeff_D-N}), respectively.\label{fig:numerics-D+N-cases}}}
\end{figure}

The results for Dirichlet and Neumann boundary conditions are shown in
Fig.~\ref{fig:numerics-D+N-cases}. All energies are divided by
$\cE_\text{PFA}$, given in Eq.~\eqref{eq:pfa-spheres}, with the
corresponding amplitude $\Phi_0^+$ (repulsive at small separations) or
$\Phi_0^-$ (attractive at small separations).  For like boundary
conditions, either Dirichlet or Neumann, the interaction is attractive
at all separations, but  for unlike boundary
conditions it is repulsive.  At large separations the numerical
results show excellent
agreement with the asymptotic expansion derived above.  Note that the
reduction of the energy compared to the PFA estimate at large distances 
depends strongly on the boundary conditions, showing the
different power laws at asymptotically large separations. In the limit
of a vanishing surface-to-surface distance ($R/d\to 1/2$), the energy
approaches the PFA estimate in all cases. Generically, the PFA
overestimates the energy:  $\cE_\text{PFA}$ is approached from
below for $R/d\to 1/2$, except in the case of  Dirichlet boundary
conditions on both spheres, where the PFA underestimates the
actual energy in a range of $0.3 \lesssim R/d < 1/2$. The deviations
from the PFA are most pronounced for Neumann boundary conditions. At a
surface-to-surface distance of $L=3R$ ($R/d=0.2$){, the} PFA
overestimates the energy by a factor of $100$.

\begin{figure}[ht]
\includegraphics[scale=0.17]{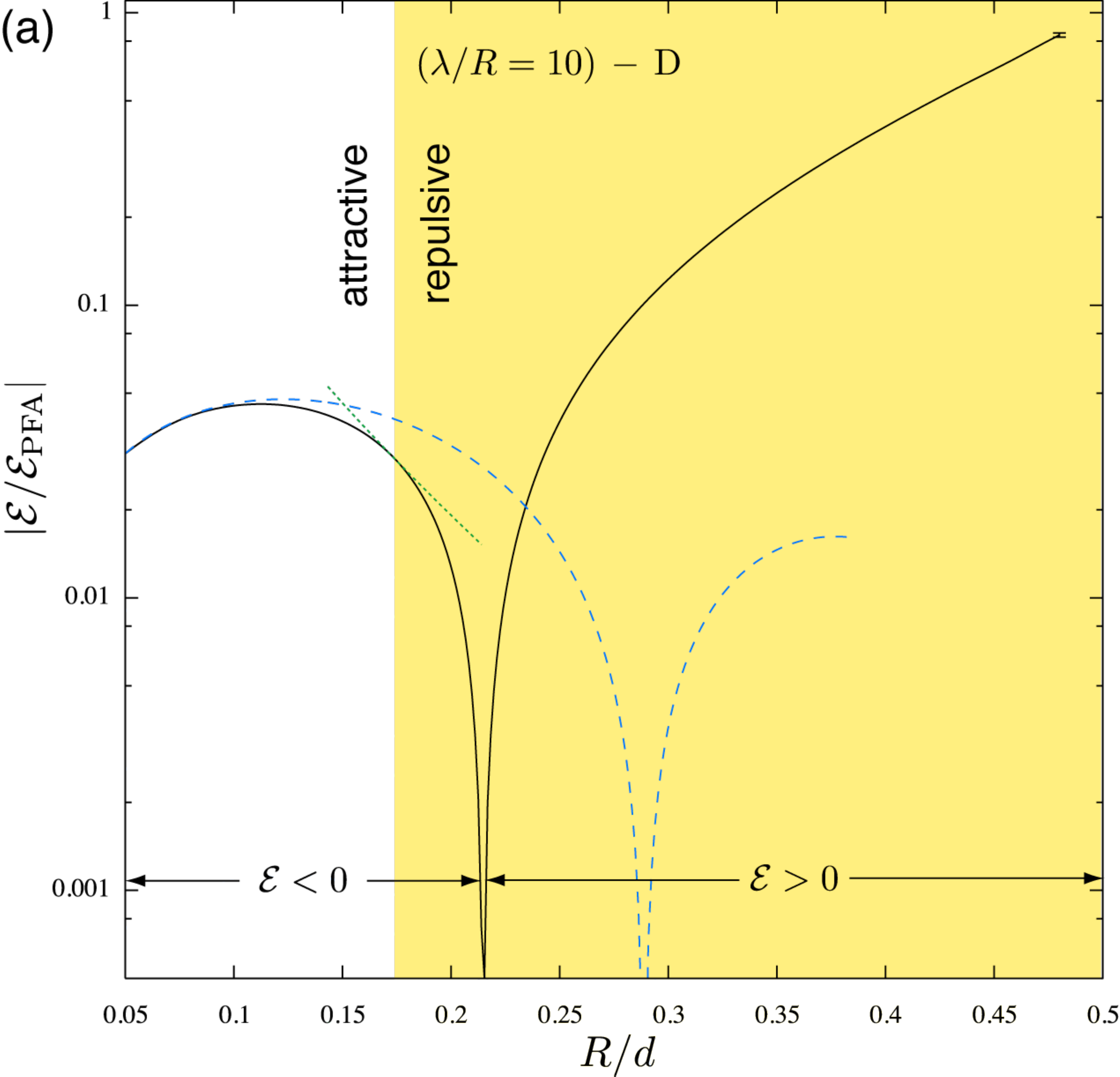}
\includegraphics[scale=0.35]{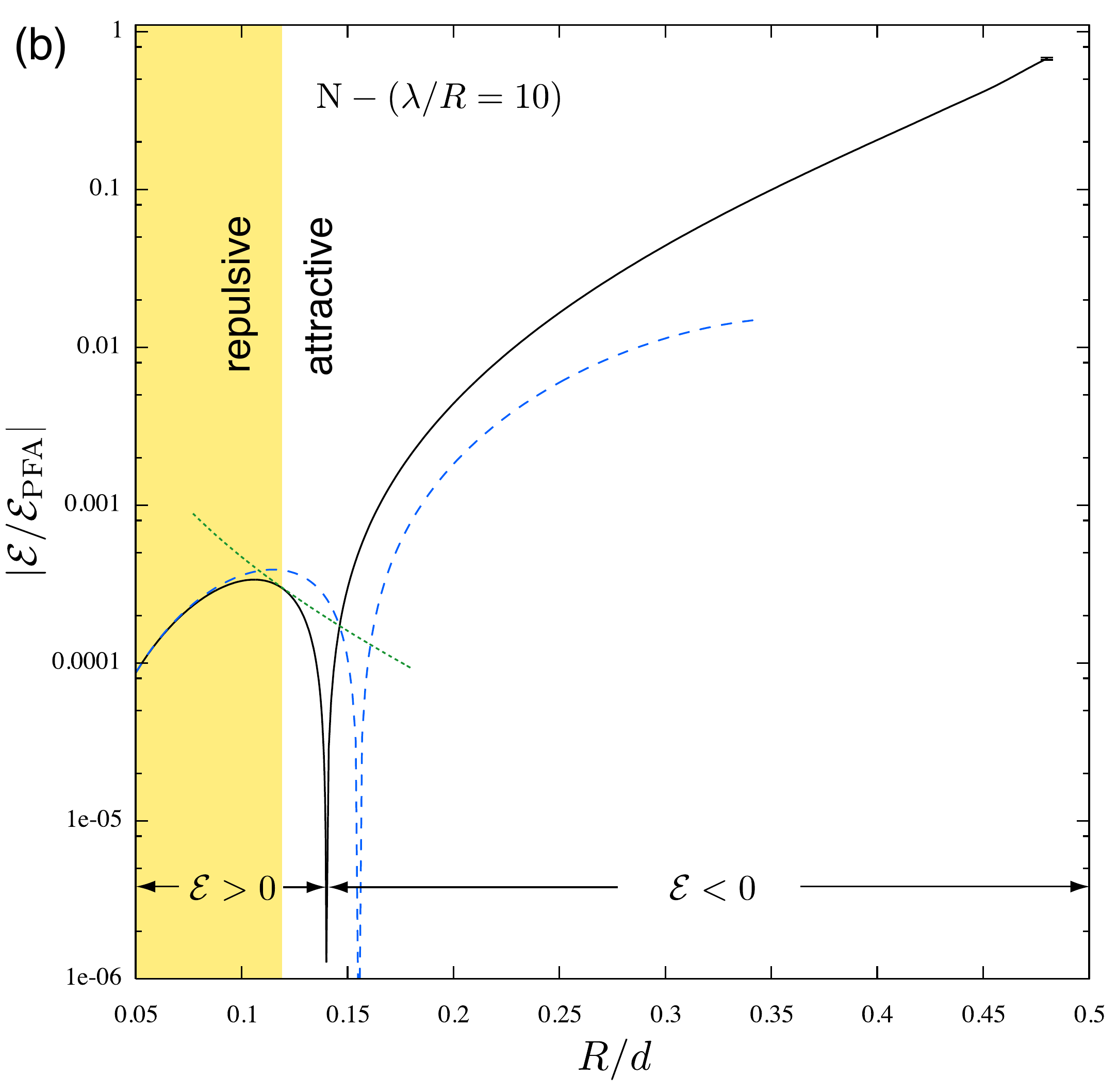}
\includegraphics[scale=0.17]{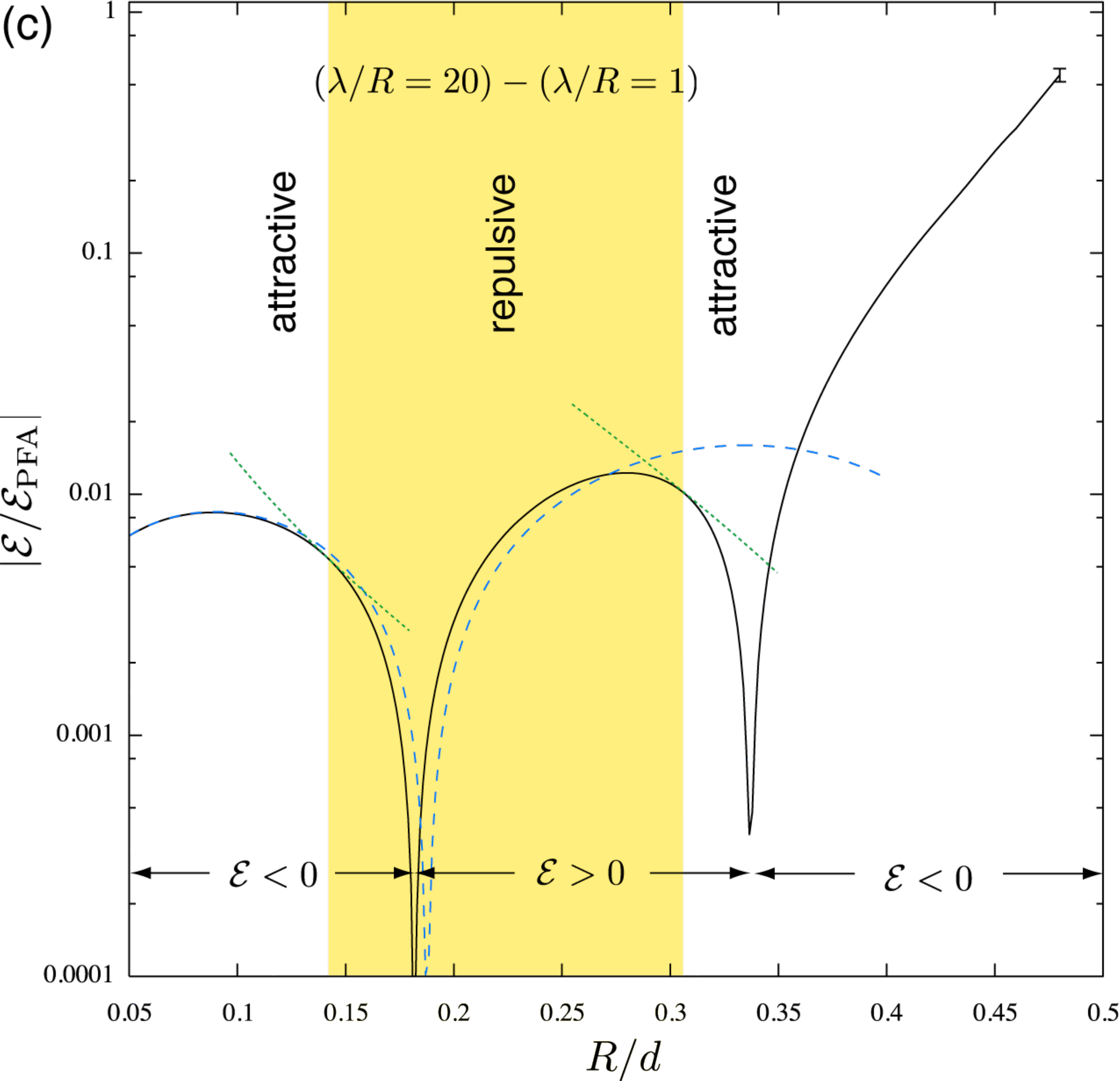}
\hspace*{-2cm}
\parbox[b]{0.62\linewidth}{\caption{The Casimir energy for two spheres
    with different Robin boundary conditions for finite
    $\lambda_\alpha$: (a) Dirichlet boundary conditions and
    $\lambda/R=10$, (b) Neumann boundary conditions and
    $\lambda/R=10$, (c) $\lambda_1/R=10$ and $\lambda_2/R=1$.  The
    solid curves correspond to extrapolated results for $l\to\infty$,
    and the dashed curves represent the asymptotic large distance
    expansion given in Eq.~(\ref{eq:spheres-energy-large-d}) with the
    coefficients given in Ref.~\cite{Emig+07}. For logarithmic
    plotting, the modulus of the energy is shown, and the sign of the
    energy is indicated at the bottom. The range of separations with a
    repulsive force is shaded. The points of vanishing force occur
    where an auxiliary function (dotted curves) is tangent to the
    solid curve, see text for details.\label{fig:numerics-finite-lambda}}}
\end{figure}

Casimir interactions for Robin boundary conditions with finite
$\lambda_\alpha$ are shown in
Fig.~\ref{fig:numerics-finite-lambda}. If $\lambda_1=\lambda_2$ the
interaction is always attractive. If the $\lambda_\alpha$ are not
equal and their ratio is sufficiently large, the Casimir force changes
sign either once or twice.  This behavior resembles the interaction of
two plates with Robin boundary conditions.  However, the criterion for
the existence of sign changes in the force now depends not only on
$\lambda_1/\lambda_2$, but on both quantities $\lambda_1/R$ and
$\lambda_2/R$ separately.  Even with $\lambda_1/\lambda_2$ fixed, for
smaller $\lambda_\alpha/R$ there can be sign changes in the force,
while for larger $\lambda_\alpha/R$ the force is attractive at all
distances. When the ratio $\lambda_1/\lambda_2$ is sufficiently large
(or formally infinite for Dirichlet or Neumann boundary conditions),
we can identify three different generic cases where sign changes in
the force occur:
\begin{itemize}
\item First, we consider Dirichlet boundary conditions ($\lambda_1=0$)
  on one sphere and a finite non-vanishing $\lambda_2/R$ at the other
  sphere. Figure~\ref{fig:numerics-finite-lambda}(a) displays the
  energy for $\lambda_2/R=10$ as a typical example. At large distances
  the energy is negative, while it is positive at short separations
  with one sign change in between. The asymptotic expansion of
  Eq.~\eqref{eq:spheres-energy-large-d} yields the exact energy at
  separations well below the sign change \cite{Emig+07}. While the
  expansion predicts qualitatively the correct overall behavior of the
  energy, it does not yield the actual position of the sign change
  correctly. Of course, for the Casimir interaction between compact
  objects, the sign of the force $\cF=-\partial \cE/\partial d$ is the
  physically important quantity, not the energy.  The distance at
  which the force vanishes cannot be deduced directly from the slope
  of the curve for $\hat\cE\equiv \cE/\cE_\text{PFA}$, since one has
\begin{equation}
  \label{eq:force-from-energy-ratio}
  \hat\cE'(d) = \frac{1}{\cE_\text{PFA}(d)} \left[
\cE'(d) +\frac{2}{d-2R} \cE(d)
\right] \, .
\end{equation}
The force vanishes at the distance $d_0$ if $\cE'(d_0)=0$, so that
\begin{equation}
\label{eq:zero-force-cond}
\hat\cE'(d_0)=\frac{2}{d_0-2R} \hat \cE(d_0) \, .
\end{equation}
Hence the distance at which the force vanishes is determined by the
position $d_0$ where the curve of the auxiliary function $t(d)=\tau
(d/R-2)^2$ is tangent to the curve of $\hat \cE$. The two unknown
quantities $d_0$ and $\tau$ are then determined by the conditions
$\hat \cE(d_0)=t(d_0)$ and $\hat \cE'(d_0)=t'(d_0)$.  This procedure
allows us to obtain the distance at which the force vanishes easily{,}
without computing derivatives numerically. The tangent segment of the
curve for $t(d)$ is shown in Fig.~\ref{fig:numerics-finite-lambda}(a)
as a dotted line.  From this construction we find that at a distance
$d_{-\Rightarrow +}$ the force changes from attractive to repulsive
for decreasing separations.  The position $d_{-\Rightarrow +}$
corresponds to a minimum of the energy and decreases with decreasing
$\lambda_2/R$, so that in the limit $\lambda_2/R\to 0$ it approaches
the case of two spheres with Dirichlet boundary conditions, where the
force is always attractive.

\item Second, we study Neumann boundary conditions on one sphere and a
  finite non-vanishing $\lambda_2/R$ at the other sphere.  As an
  example we choose again $\lambda_2/R=10$, as shown in
  Fig.~\ref{fig:numerics-finite-lambda}(b). The energy is positive at
  large distances and becomes negative at small distances.  The
  asymptotic expansion is found to be valid well below the separation
  where the sign of the energy changes \cite{Emig+07}.  Hence, the
  expansion describes the behavior of the energy qualitatively, but
  does not predict the precise position of the sign change. The sign
  change of the force can be obtained by the method described above.
  At a position $d_{+\Rightarrow -}$, the force changes from repulsive
  to attractive with decreasing separation and the energy is maximal.
  A decreasing (increasing) $\lambda_2/R$ shifts $d_{+\Rightarrow -}$
  to smaller (larger) separations.  This result is consistent with an
  entirely repulsive (attractive) force for Neumann-Dirichlet
  (Neumann-Neumann) boundary conditions.

\item
 The third case is obtained if both $\lambda_\alpha$ are finite
  and non-zero. A typical example with $\lambda_1/R=20$ and
  $\lambda_2/R=1$ is shown in
  Fig.~\ref{fig:numerics-finite-lambda}(c). The energy is negative
  both at large and small separations but turns positive at
  intermediate distances.  The asymptotic expansion applies again at
  sufficiently large separations beyond the position where the energy
  becomes positive. For values of the ratio $\lambda_1/\lambda_2$ that
  are larger than an $R$-dependent threshold, the force changes sign
  twice, so that it is repulsive between the separations
  $d_{-\Rightarrow +}$ and $d_{+\Rightarrow -}$.  The energy has a
  minimum (maximum) at $d_{-\Rightarrow +}$ ($d_{+\Rightarrow -}$).
  If $\lambda_1/R$ increases and $\lambda_2/R$ decreases, the
  repulsive region grows until eventually the force becomes repulsive
  at all separations, corresponding to the limit of Dirichlet/Neumann
  boundary conditions.  Decreasing $\lambda_1/R$ and increasing
  $\lambda_2/R$ reduces the interval with repulsion.  In this case,
  first the zeros of the energy disappear, leaving negative energy at
  all distances but still a repulsive region, and then the two
  positions where the force vanishes merge, leaving an entirely
  attractive force.
\end{itemize}

\section{Applications: Electromagnetic field}
\label{sec:appl-em}

As a specific example for the electromagnetic field, we consider two
identical dielectric spheres.  Due to symmetry, the multipoles are
decoupled so that the T-matrix is diagonal,
\begin{equation}
\label{eq:t-matrix-elem-sphere}
  T^{11}_{lmlm}=(-1)^l \frac{\pi}{2} \frac{\eta I_{l+{1\over 2}}(z)
\left[I_{l+{1\over 2}}(nz)+2nzI'_{l+{1\over 2}}(nz)\right] - n I_{l+{1\over 2}}(nz)
\left[I_{l+{1\over 2}}(z)+2z I'_{l+{1\over 2}}(z)\right]}
{\eta K_{l+{1\over 2}}(z)
\left[I_{l+{1\over 2}}(nz)+2nzI'_{l+{1\over 2}}(nz)\right] - n I_{l+{1\over 2}}(nz)
\left[K_{l+{1\over 2}}(z)+2z K'_{l+{1\over 2}}(z)\right]} \, ,
\end{equation}
where the sphere radius is $R$, $z=\kappa R$,
$n=\sqrt{\epsilon(i\kappa)\mu(i\kappa)}$,
$\eta=\sqrt{\epsilon(i\kappa)/\mu(i\kappa)}$, and $I_{l+{1\over 2}}$,
$K_{l+{1\over 2}}$ are Bessel functions.  $T^{22}_{lmlm}$ is obtained
from Eq.~(\ref{eq:t-matrix-elem-sphere}) by interchanging $\epsilon$
and $\mu$. For all partial waves, the {\it leading} low frequency
contribution is determined by the {\it static} electric 
multipole polarizability, $\alpha^\textsc{E}_l =
[(\epsilon-1)/(\epsilon+(l+1)/l)]R^{2l+1}$,
and the corresponding magnetic polarizability, $\alpha^\textsc{M}_l =
[(\mu-1)/(\mu+(l+1)/l)]R^{2l+1}$.  Including the next to leading terms,
the T-matrix has the structure
\begin{equation}
  \label{eq:T-low-kappa}
  T^{11}_{lmlm}=\kappa^{2l}\bigg[\frac{(-1)^{l-1}(l+1)\alpha_l^\textsc{M}}{l (2l+1)!! (2l-1)!!}  \kappa 
+ \gamma^\textsc{M}_{l3}\kappa^{3}+\gamma^\textsc{M}_{l4}\kappa^{4} +\ldots\bigg] \, ,
\nonumber
\end{equation}
and $T^{22}_{lmlm}$ is obtained by $\alpha^\textsc{M}_l \to
\alpha^\textsc{E}_l$,
$\gamma_{ln}^\textsc{M}\to\gamma^\textsc{E}_{ln}$.  The first terms
are
$\gamma^\textsc{M}_{13}=-[4+\mu(\epsilon\mu+\mu-6)]/[5(\mu+2)^2]R^5$,
$\gamma^\textsc{M}_{14}=(4/9)[(\mu-1)/(\mu+2)]^2R^6$, and
$\gamma^\textsc{E}_{13}$, $\gamma^\textsc{E}_{14}$ are obtained again
by the replacement, $\mu\to\epsilon$. Now we can apply our general
formula in Eq.~\eqref{finalform2} (with a factor of $1/2$ and the
translation matrices for the electromagnetic field \cite{wittmann}) to two dielectric
spheres with center-to-center distance $d$. For simplicity, we
restrict to two partial waves ($l=2$) and two scatterings ($p=1$),
which yields the exact Casimir energy to order $d^{-10}$. Matrix
operations are performed with {\tt Mathematica}, and we find the
interaction
\begin{eqnarray}
  \label{eq:2-dielectric-spheres}
&&  \cE=-\frac{\hbar c}{\pi} \bigg\{ \bigg[ \frac{23}{4} \left((\alpha^\textsc{E}_1)^2+
(\alpha^\textsc{M}_1)^2\right) - \frac{7}{2} \alpha^\textsc{E}_1\alpha^\textsc{M}_1 \bigg] \frac{1}{d^7} \nonumber\\
& &+ \frac{9}{16} \big[ \alpha^\textsc{E}_1 \big(59 \alpha^\textsc{E}_2 -11
\alpha^\textsc{M}_2+86 \gamma^\textsc{E}_{13}  -54 \gamma^\textsc{M}_{13} \big)
+ \, \textsc{e} \leftrightarrow \textsc{m} \,  \big] \frac{1}{d^9} \nonumber\\
& &+ \frac{315}{16} \big[  \alpha^\textsc{E}_1 \big( 7 \gamma^\textsc{E}_{14} - 5 \gamma^\textsc{M}_{14} \big)
+ \, \textsc{e} \leftrightarrow \textsc{m} \,
\big] \frac{1}{d^{10}} + \dots \bigg\} \, ,
\end{eqnarray}
where $ \textsc{E} \leftrightarrow \textsc{M}$ indicates 
terms with exchanged superscripts. The
leading term, $\sim d^{-7}$, has precisely the form of the
Casimir-Polder force between two atoms \cite{Casimir+48}, including
magnetic effects \cite{Feinberg+70}. 
The higher order terms are new, and provide the first
systematic result for dielectrics with strong curvature. 
There is no $\sim 1/d^8$ term. 

\begin{figure}[ht]
\begin{center}
\includegraphics[width=.7\linewidth]{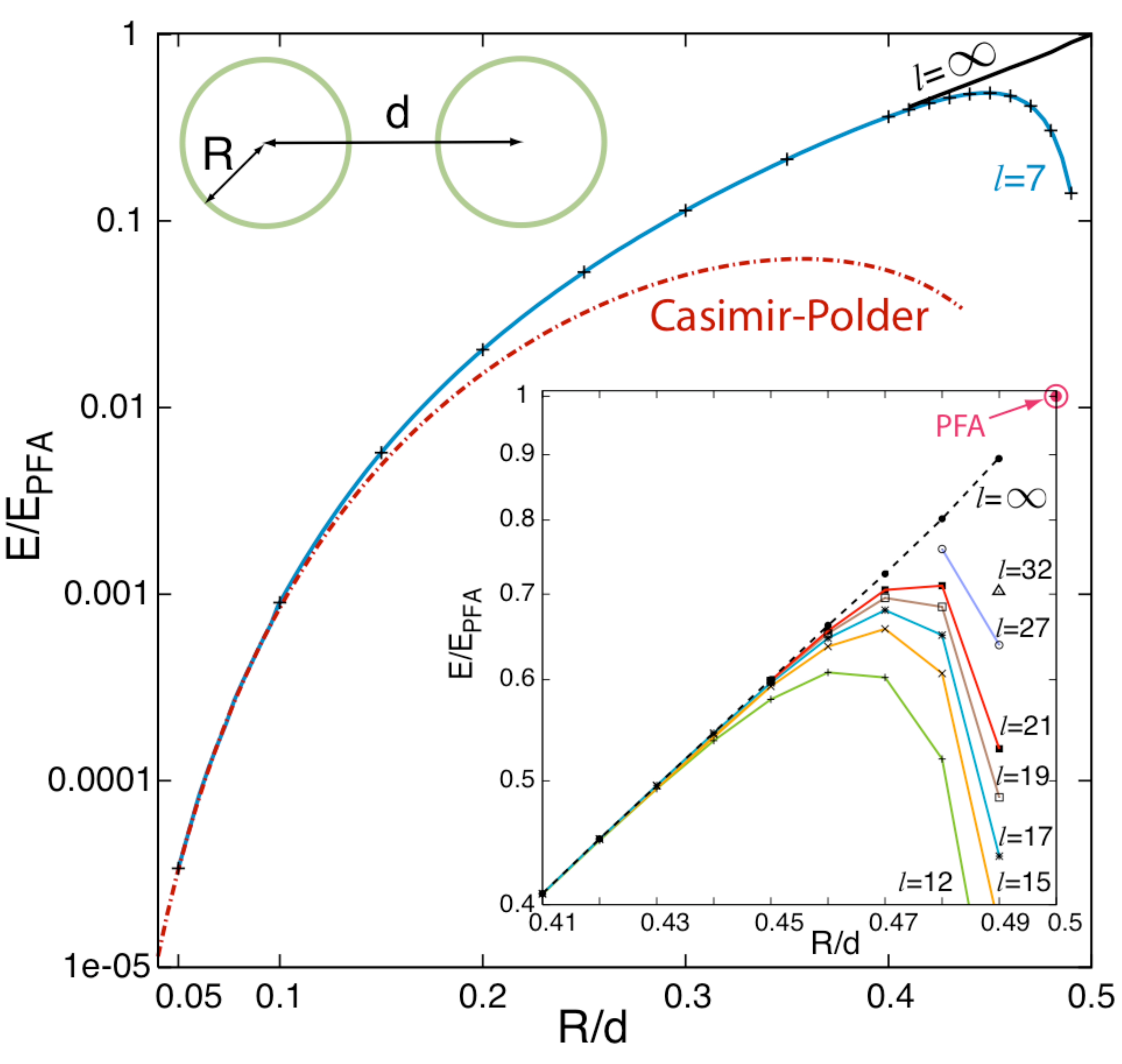}
\caption{\label{fig:energy}Casimir energy of two metal spheres,
  divided by the PFA estimate $\cE_\text{PFA}=-(\pi^3/1440)\hbar c
  R/(d-2R)^2$, which holds only in the limit $R/d\to 1/2$. The label
  $l$ denotes the multipole order of truncation. The curves $l=\infty$
  are obtained by extrapolation. The Casimir-Polder curve is the
  leading term of Eq.~\eqref{eq:2-metal-spheres}. Inset: Convergence
  at short separations.}
\end{center}
\end{figure}

The limit of perfect metals follows for $\epsilon\to\infty$, $\mu\to
0$.  Then higher orders are easily included, yielding an asymptotic
series
\begin{equation}
  \label{eq:2-metal-spheres}
  \cE=-\frac{\hbar c}{\pi} \frac{R^6}{d^7} \sum_{n=0}^\infty c_n \left(\frac{R}{d}\right)^n \, ,
\end{equation}
where the first 10 coefficients are $c_0=143/16$,
$c_1=0$, $c_2=7947/160$,
$c_3=2065/32$, $c_4=27705347/100800$,
$c_5=-55251/64$, $c_6=1373212550401/144506880$, $c_7=-7583389/320$,
$c_8=-2516749144274023/44508119040$, $c_9=274953589659739/275251200$.
This series is obtained by expanding in powers of $\mathbb{N}$ and
frequency $\kappa$, and does not converge for any fixed $R/d$.  To
obtain the energy at all separations, one has to compute
Eq.~\eqref{finalform2} without these expansions. This is done
as before in the case of a scalar field: We
truncate the matrix $\bbN$ at a finite multipole order $l$, and
compute the determinant and the integral numerically. The result is
shown in Fig.~\ref{fig:energy} for perfect metal spheres. Our data
indicate again that the energy converges as $e^{-\delta (d/R-2) l}$ to its
exact value at $\l \to \infty$, with $\delta\sim{\cal O}(1)$.  Our
result spans all separations between the Casimir-Polder limit for
$d\gg R$, and the proximity force approximation (PFA) for $R/d \to
1/2$.  At a surface-to-surface distance $L = 4R/3$ ($R/d=0.3$), PFA
overestimates the energy by a factor of 10. Including up to $l=32$ and
extrapolating based on the exponential fit, we can accurately
determine the Casimir energy down to $R/d=0.49$, {\it i.e.\/} $L=0.04
R$.  A similar numerical evaluation can be also applied to dielectrics
\cite{Emig+07}.

\ack This work was supported by the U.~S.~Department of Energy (DOE)
under cooperative research agreement \#DF-FC02-94ER40818 (RLJ), and by
a Heisenberg Fellowship from the German Research Foundation (TE).

\end{document}